\documentclass[apj]{emulateapj}
\usepackage{longtable}

\newcommand{\cena}{NGC~5128}
\newcommand{\etal}{et~al.\ }

\newcommand{\Hb}{H$\beta$}       
\newcommand{\kms}{km~s$^{-1}$}

\newcommand{\uv}{$U$--$V$}
\newcommand{\vi}{$V$--$I$}
\newcommand{\ubvri}{{\it UBVRI}}

\journalinfo{Astrophysical Journal Supplement Series, 150, 2004 February}
\submitted{Received 2003 August 14; accepted 2003 November 3}  

\shorttitle{Globular Cluster System of NGC 5128: Survey and Catalogs}
\shortauthors{Peng, Ford, \& Freeman}

\begin{document}


\title{The Globular Cluster System of NGC 5128 I. Survey and Catalogs}


\author{Eric W.\ Peng\altaffilmark{1,2}, Holland C.\ Ford\altaffilmark{1,3}}
\affil{Department of Physics and Astronomy, Johns Hopkins
        University, Baltimore, MD, 21218, USA}
\email{ericpeng@pha.jhu.edu, ford@pha.jhu.edu}

\and

\author{Kenneth C.\ Freeman}
\affil{RSAA, Australian National University, Canberra, ACT, Australia}
\email{kcf@mso.anu.edu.au}


\altaffiltext{1}{Visiting Astronomer, Cerro Tololo Inter-American Observatory,
which is operated by the Association of Universities for Research in
Astronomy, Inc.\ (AURA) under cooperative agreement with the National Science
Foundation.}
\altaffiltext{2}{Current address: 136 Frelinghuysen Road, Physics and
Astronomy, Rutgers University, Piscataway, NJ 08854, USA;
ericpeng@physics.rutgers.edu}
\altaffiltext{3}{Space Telescope Science Institute, 3700 San Martin Drive,
        Baltimore, MD 21218, USA}


\begin{abstract}
We present the results of a photometric and
spectroscopic survey of the globular cluster system of \cena\
(Centaurus~A), a galaxy whose proximity makes it an important target for
early-type galaxy studies.  
We imaged three fields in {\it UBVRI} that extend 50 and 30~kpc along
the major and minor axes, respectively.  We used both color and size
information to develop efficient selection criteria for differentiating
between star clusters and foreground stars.  In total, we obtained new
velocities for 138 globular clusters, nearly tripling the number of
known clusters, and bringing the confirmed total in
\cena\ to 215.  We present a full catalog of all known GCs, with their
positions, photometry, and velocities.  In addition, we 
present catalogs of other objects observed, such as foreground stars,
background galaxies, three Galactic white dwarfs, seven background QSOs,
and 52 optical counterparts to known X-ray point sources.
We also report an observation of the cluster G169,
in which we confirm the existence of a bright emission line object.
This object, however, is unlikely to be a planetary nebula, but may be a
supernova remnant.
\end{abstract}




\section{Introduction}

Systems of globular clusters are a nearly ubiquitous feature of all
nearby galaxies.  Globular clusters (GCs), which are typically old with
sub-solar metallicities, are the most visible remnants of intense star
formation that occurred in a galaxy's distant past.  As single-age,
single-metallicity stellar populations, GCs are ideal for studying
the fossil remains of a galaxy's star formation and metal-enrichment
history.

The past decade has seen a rapid growth in the study of extragalactic GC
systems.  The availability of the Hubble Space Telescope (HST) in
particular has enabled the study of GC systems in galaxies well beyond
the Local Group.  One of the more striking results of these studies is
the frequency with which the metallicity distributions of these GC
systems are bimodal (e.g. Larsen \etal 2001; Kundu \& Whitmore 2001).
Different scenarios of galaxy formation (or at least of globular cluster
system formation) have been proposed or adapted to explain the observed
metallicity distributions and other properties of GC systems:
mergers of spiral galaxies (Ashman \& Zepf 1992), multiple {\it in situ}
star formation epochs (Forbes, Brodie, \& Grillmair 1997),
dissipationless hierarchical merging of protogalactic clumps (C\^{o}t\'{e},
Marzke, \& West 1998), and more generalized hierarchical merging
(Beasley \etal 2002).  The question that all of these scenarios address
largely concerns the nature and time frame of merging or gas dissipation.  
Elements of all of these scenarios are supported by various studies of
different facets of galaxy formation and evolution.  For
example, the supernova winds that some use to explain the suppression of
star formation in dwarf galaxies at high redshift (Dekel \& Silk 1986)
has also been proposed as a possible mechanism for the truncation of
metal-poor GC formation at early times (Beasley \etal 2002).  Similarly,
hierarchical merging models are often used to explain the mass assembly of
present-day galaxies.

While most spheroid formation may have occurred in the past
(at redshifts $z > 2$), present-day examples of recent merger remnants
may give us a window onto this distant epoch.
Locally, there is compelling evidence that some
ellipticals have recently interacted or merged with another galaxy.
Many ellipticals possess large-scale disks of gas and dust.  Faint
structure in the form of loops, shells, ripples, and tails are
especially visible in the outer regions of ellipticals (Malin \&
Carter 1983), and are presumably the aftermaths of a recent
interaction.
One such galaxy, and perhaps the best candidate for study, is the nearby
elliptical, \cena.  We have chosen to examine of the stellar content of
\cena\ in order to further elucidate its formation history and gain
insight on the formation of other ellipticals.  Our survey of the
planetary nebula (PN) system in \cena's outer halo represents the kinematics
of the field star population, and is presented in a separate
paper (Peng, Ford, \& Freeman 2004).  In this paper, we describe our
photometric and spectroscopic survey for globular clusters, and present
the resulting catalogs of objects.

\vspace{0.3cm}

\section{The GC system of \cena}

As the nearest large elliptical galaxy, and as a recent merger remnant, 
\cena\ (also known as the radio source 
Centaurus A) is an obvious target for GC system studies.  
At a distance of 3.5 Mpc (Hui \etal 1993), \cena\ is the only
early-type member of the Centaurus group, an environment of lower
density than galaxy clusters which harbor many of the luminous 
ellipticals previously studied.  There is also much observational
evidence that \cena\
has experienced one or more major merging events, including a
warped disk of gas and dust at its center, faint shells and
extensions in its light profile (Malin 1978), and a young tidal stream
in its halo (Peng, Ford, Freeman, \& White 2002).  It is also known to
have a bimodal distribution of globular cluster metallicities (Zepf \&
Ashman 1993; Held \etal 1997).  \cena\ is by far the nearest active
radio galaxy, and exhibits signatures of recent star formation where
the radio jet has interacted with shells of \ion{H}{1} (Graham 1998).
For a recent and complete review of this galaxy, see Israel (1998).
The combination of its proximity and post-merger state makes \cena\ an
excellent target for a detailed study.

\cena's peculiar appearance and nature has long led astronomers to believe
that it is somehow unique.  However, as Ebneter \& Balick (1983)
point out in their review, the galaxy's proximity permits us to
collect data more detailed than for most other galaxies, and thus makes
it seem more peculiar. In fact, \cena\ is rather typical member of the
population of dusty elliptical galaxies and radio galaxies.  Massive
galaxies with old stellar populations and central dust obscuration are
known to host radio sources in both the local universe and at redshifts
out to and beyond $z\sim1$ (e.g.\ Zirm, Dickinson, \& Dey 2003).

Some of the early work on extragalactic GCs was done
in \cena.  Noting a slightly diffuse 17th magnitude object on
photographic plates, Graham \& Phillips (1980; GP80) obtained follow-up
spectroscopy and identified the first GC in this
galaxy.  Five more GCs were confirmed by van den
Bergh, Hesser, \& G.Harris (1981; VHH81).  Using {\it UVR} star counts,
G.Harris, Hesser, H.Harris, \& Curry (1984) estimated that the total
cluster population in \cena\ is 1200--1900.   Subsequent spectroscopic
work (Hesser, H.Harris, \& G.Harris 1986 [HHH86]; Sharples 1988) 
increased the number of GCs
with radial velocities.  Finally, G.Harris \etal (1992; HGHH92) obtained CCD
photometry on the Washington system for 62 confirmed GCs.  This study
produced one of the best GC system metallicity distributions at the
time, and was later cited by Zepf \& Ashman (1993) as evidence for
bimodality.  

Since then, the relatively small fields of view of modern detectors
have forced studies to concentrated on 
the inner regions of the \cena\ GC system.  Minniti \etal (1996; MAGJM96) used
infrared imaging to identify possible intermediate-age, metal-rich GCs.
Holland, C\^{o}t\'{e}, \& Hesser (1999; HCH99) used HST WFPC2
imaging to identify spatially resolved GC candidates in the region 
near the dust lane.  Rejkuba (2001) showed what is possible with
excellent (0\farcs6) seeing and 8-meter telescopes when she used
imaging taken at the Very Large Telescope to define a sample of fainter
GC candidates based on their resolved appearance in two 
$7\arcmin\times7\arcmin$ fields.  Only with the availability of
mosaic cameras are CCD studies now able to cover an area
comparable to the photographic studies of the 1980s.  Recent
programs include this one and an imaging study using 
the Big Throughput Camera on the CTIO 4-meter (G.Harris \& W.Harris
2003) with the Washington photometric system (Canterna 1976).

Unfortunately, \cena's nearness and relative proximity to the Galactic
bulge and disk ($l=309\fdg5$, $b=+19.4$) conspire to make GC studies
difficult.  At the distance of \cena, 
$1\arcmin = 1.02$~kpc, which means that \cena's
halo subtends over two degrees of sky.  The density of foreground stars is
also high, such that even in the central regions, all but a few percent
of objects within the magnitude range expected for \cena\ GCs
are actually stars in our own Milky Way.  Spectroscopic
follow-up to obtain radial velocities is the only way to confidently 
identify true GCs.  It is only with the relatively 
recent development of wide-field
imaging and spectroscopy that we have been able to substantially
increase the sample of known GCs in \cena.  Despite these difficulties,
the return on these investigations has the potential to be large ---
\cena\ is the only large elliptical whose stellar
populations can be studied in detail, and its proximity makes it the
local benchmark for studies of more distant early-type galaxies.

\section{A {\it UBVRI} Broadband \\ Imaging Survey of NGC 5128}
\subsection{Observations}

\begin{deluxetable*}{cccllllll}
\tabletypesize{\scriptsize}
\tablecaption{CTIO Mosaic {\it UBVRI} Observations \label{table:mosaicobs}}
\tablewidth{0pt}
\tablehead{
\colhead{Field} & \colhead{RA} & \colhead{DEC} &
\colhead{Field-of-View} & 
\colhead{U} &
\colhead{B} &
\colhead{V} &
\colhead{R} &
\colhead{I} \\
\colhead{} & \colhead{(J2000)} & \colhead{J2000} & \colhead{arcmin} & 
\colhead{(sec)} & \colhead{(sec)} & \colhead{(sec)} & 
\colhead{(sec)} & \colhead{(sec)} \\
\colhead{} & \colhead{} & \colhead{} & \colhead{} & 
\colhead{(\arcsec)$^{a}$} & \colhead{(\arcsec)$^{a}$} & 
\colhead{(\arcsec)$^{a}$} & \colhead{(\arcsec)$^{a}$} & 
\colhead{(\arcsec)$^{a}$} 
}
\startdata
CTR & 13:25:30 & $-$43:01:33 & $37.6\times38.6$ & 3600 & 1500 & 1800 &
1000 & 1000 \\
 & & & & 1.20 & 1.27 & 1.25 & 1.24 & 1.18 \\
NE & 13:26:46 & $-$42:28:25 & $36.0\times38.1$ & $3719^{b}$ & 1500 & 
1800 & 1800 & 1500 \\
 & & & & 1.38 & 1.32 & 1.21 & 1.04 & 1.26 \\
S & 13:25:27 & $-$43:26:30 & $37.2\times39.1$ & $3662^{b}$ & 1500 & 
1800 & 1800 & $1608^{b}$ \\
 & & & & 1.26 & 1.12 & 0.96 & 1.11 & 1.97 \\
\enddata
\tablenotetext{a}{Average seeing (FWHM on image)}
\tablenotetext{b}{Effective total exposure time (some images were taken
during non-photometric conditions)}
\end{deluxetable*}

%
\begin{figure*}
\plotone{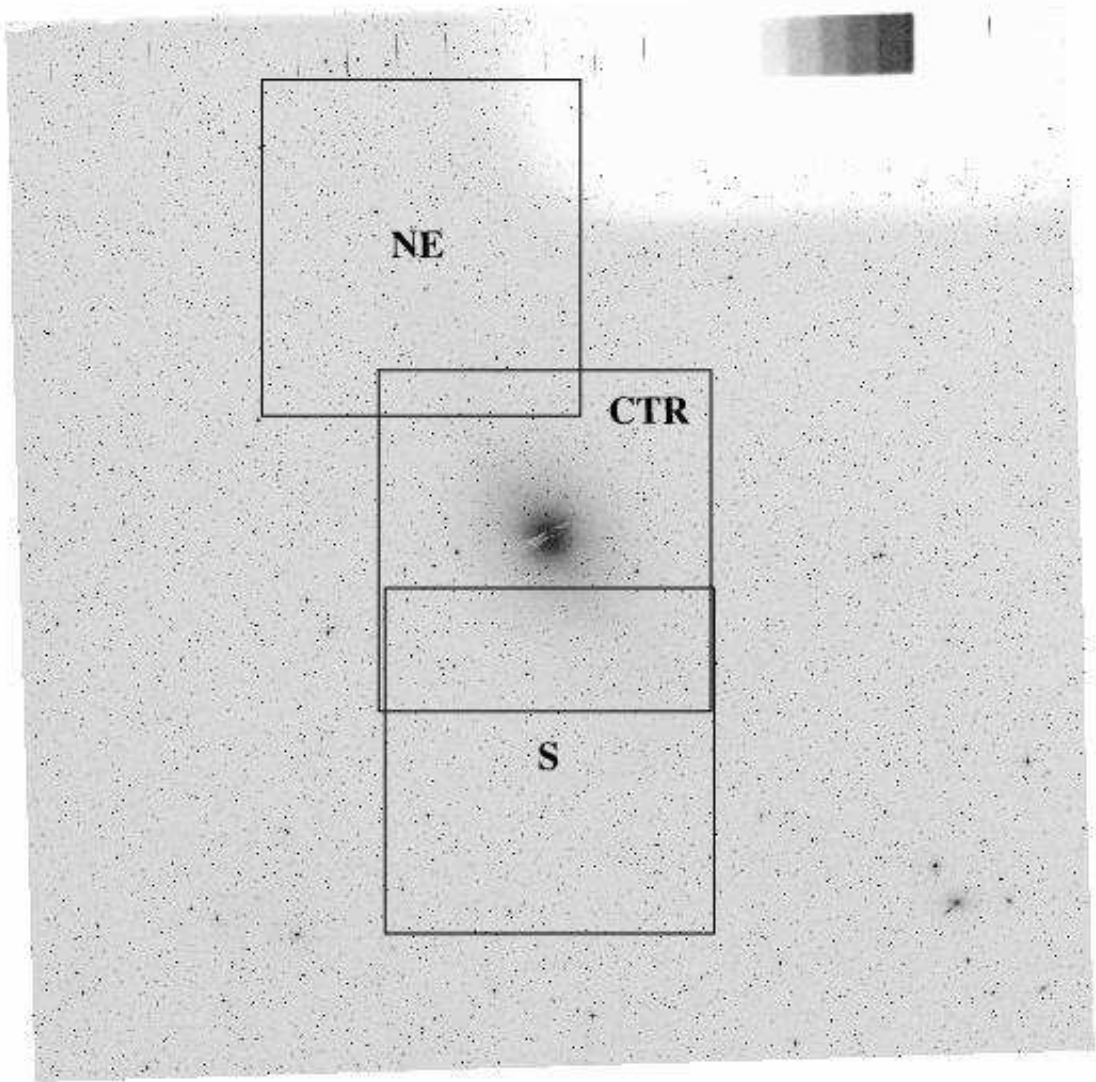}
\caption[Three Mosaic II fields overplotted on DSS image of \cena]
{Our three Mosaic II fields overplotted on a
$2\arcdeg\times2\arcdeg$ Digitized Sky Survey image of \cena.  North is up and
East is to the left.  Fields are referred to in the text by the
positions relative to the galaxy: CTR, NE, and S.  The exact centers and
dimensions of each field is presented in Table~\ref{table:mosaicobs}.
\label{figure:gcfields}}
\end{figure*}

\begin{figure*}
\plotone{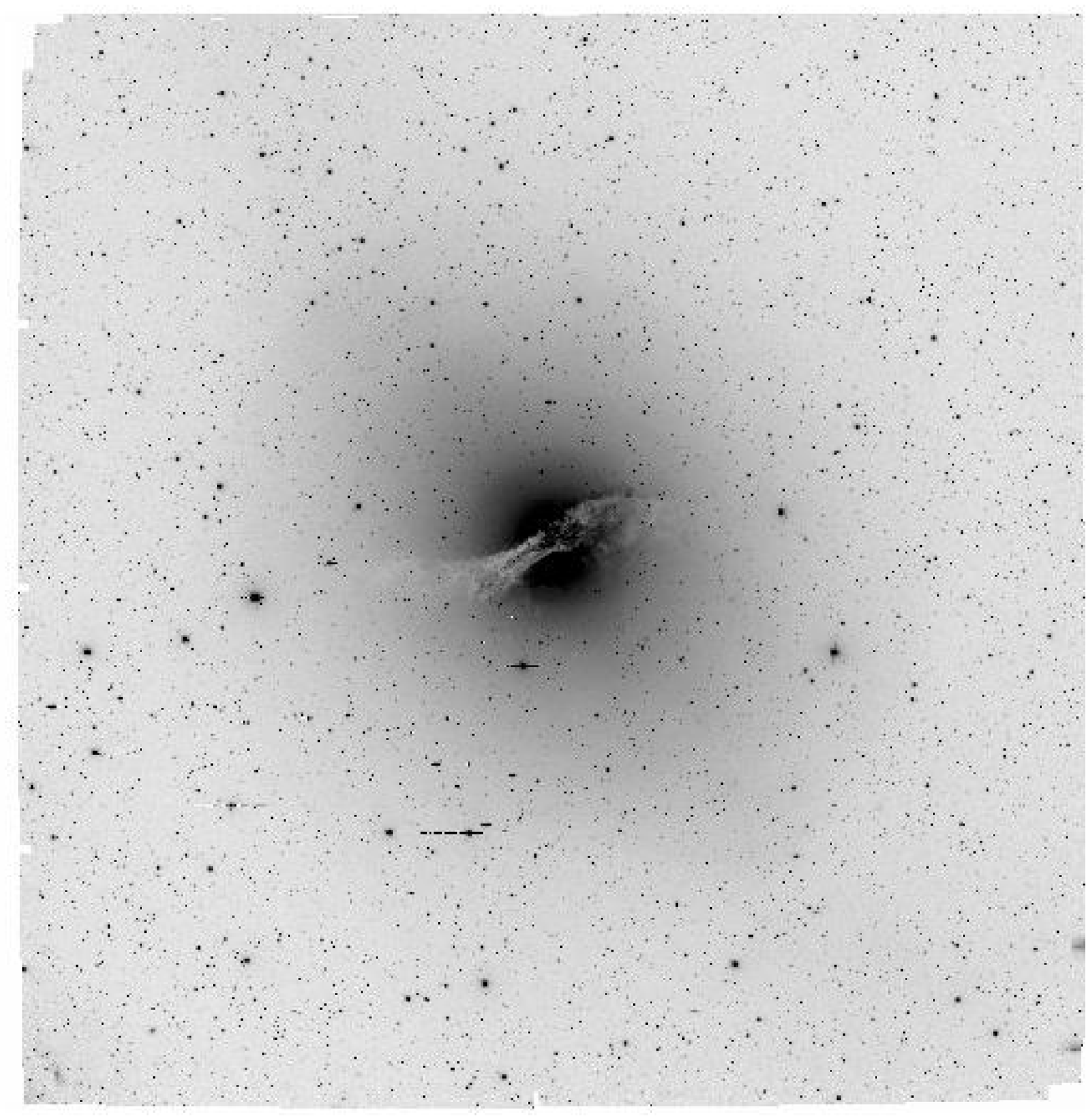}
\caption[{\it V}-band image of the CTR field]
{{\it V}-band image of the CTR field.  North is up, East is to
the left. \label{figure:GCctrV}}
\end{figure*}

\begin{figure*}
\plotone{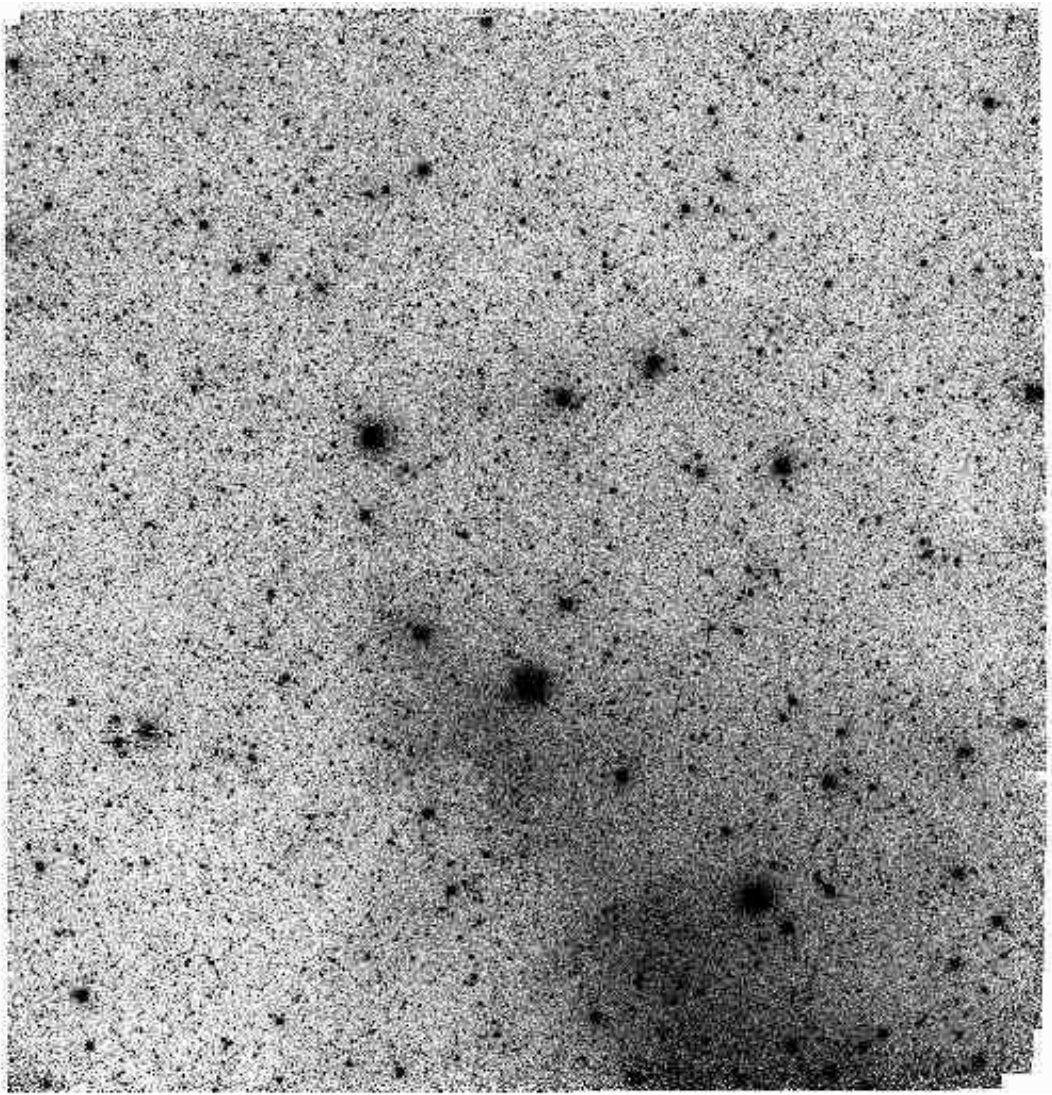}
\caption[{\it V}-band image of the NE field]
{{\it V}-band image of the NE field.\label{figure:GCneV}}
\end{figure*}

\begin{figure*}
\plotone{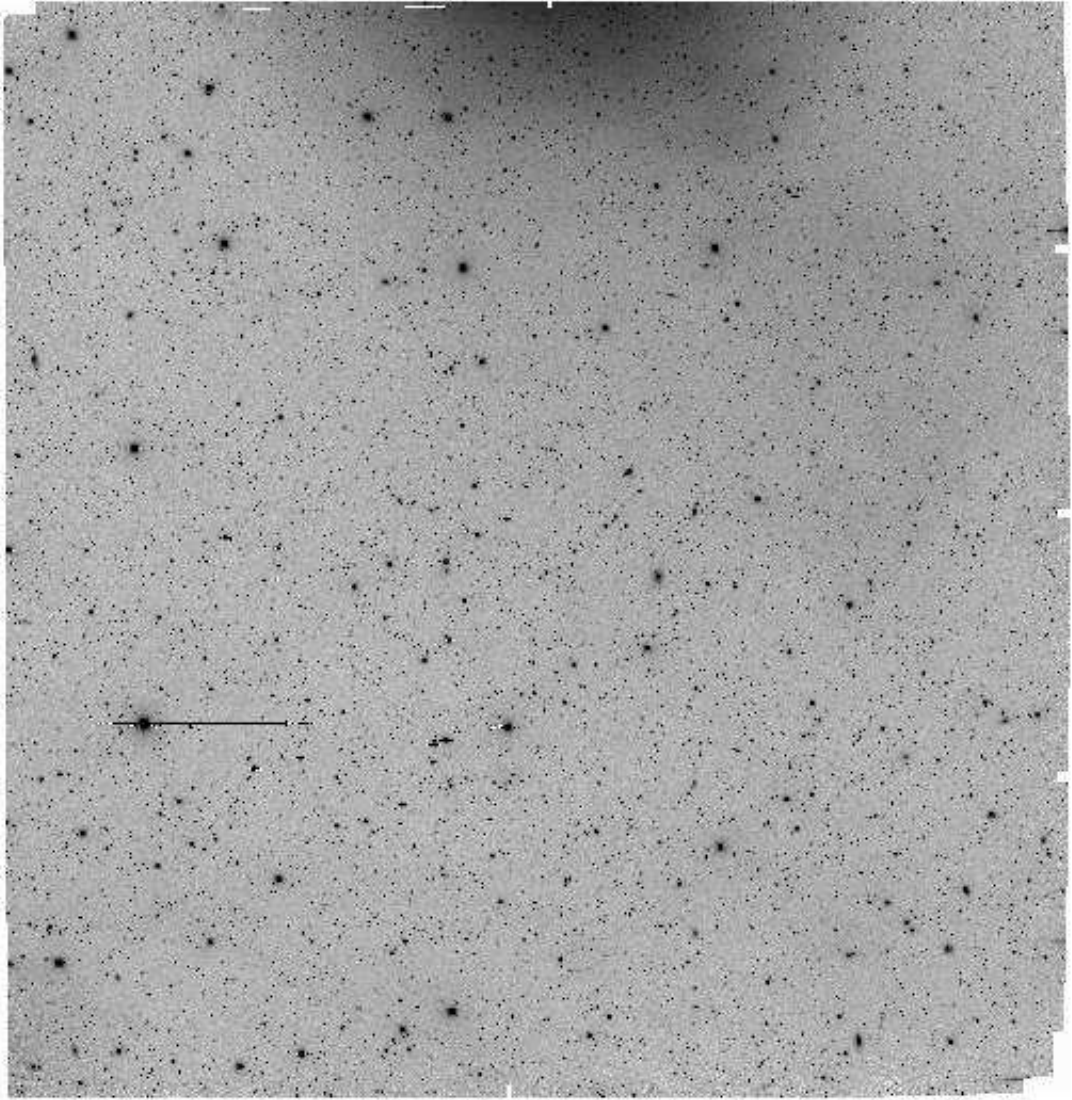}
\caption[{\it V}-band image of the S field]
{{\it V}-band image of the S field.\label{figure:GCsV}}
\end{figure*}

We conducted our imaging survey with the Mosaic II optical CCD camera on the
Blanco 4-meter telescope at Cerro Tololo Inter-American observatory
(CTIO).  We observed three fields with Mosaic II,
which has $0\farcs26$~pixels and a 0\fdg5 field-of-view.
We imaged through the Johnson-Cousins {\it
UBVRI} filters on the nights of 1--3 June 2000.  These 
38\arcmin$\times$38\arcmin\ dithered fields are shown in
Figure~\ref{figure:gcfields}.  The center field (CTR) is centered on the
galaxy.  The northeast (NE) field was chosen to follow
the faint halo light out along the major axis, 
while the south (S) field
was chosen to avoid galactic cirrus and observe a large ``sky'' area, as
determined from Malin's deep photographic image
(Malin 1978).  The {\it V}-band image of each field is
shown in Figures~\ref{figure:GCctrV}, \ref{figure:GCneV}, and
\ref{figure:GCsV}.  The total exposure times in
{\it UBVRI} were 3600s, 1500s, 1800s, 1000s, and 1000s, respectively, 
for the CTR field.  For the other two fields, we went slightly
deeper in {\it U},{\it R}, and {\it I} with exposure times of 3900s, 1500s,
1800s, 1800s, and 1500s, respectively.  We split our observations in
each band into a series of five dithered exposures (six for {\it
U} and {\it I} in the NE and S fields). Conditions were
non-photometric on the first night, but were photometric for the
remainder of the observing run.  All images taken on the first night
were calibrated by images taken on the subsequent nights.  The seeing
had a range of 1\arcsec--2\arcsec, with a median value of $1\farcs24$.  
For the image sets with some non-photometric data, we calculate the
``effective'' exposure time, which is the equivalent exposure time
that would be necessary during photometric conditions to reach the same
flux levels we measure in our images.
This information is summarized in Table~\ref{table:mosaicobs}.

\subsection{Data Reduction}

The CTIO Mosaic camera consists of eight $2048\times4096$ CCDs arranged in a
configuration that produces a $8196\times8196$ pixel focal plane.
We reduced the data using the IRAF\footnote{IRAF is distributed by the 
National Optical Astronomy Observatories, which are operated by the
Association of Universities for Research in Astronomy, Inc., under
cooperative agreement with the National Science Foundation.}
tasks provided in the MSCRED
package (Valdes 1998).  These tools were specially designed for the
reduction of mosaic CCD data.  We applied standard techniques for CCD
reduction, including overscan and bias subtraction, and flat-fielding.  We
took and used both dome, twilight, and night sky flats.  We applied a
sinc interpolation for astrometric regridding.  The varying spatial
scale in the field of view, where pixels farther from the image center
subtend more sky, means that flat-fielding corrections 
overestimate the sensitivity of pixels at the field's edge.  The
regridding process compensates for this effect and redistributes the
flux properly.

Once images were regridded, we found it necessary to introduce another
scaling correction.  Due to either gain or linearity variations, 
the sky levels as traced across CCD boundaries were not continuous,
having discontinuities of 0-3\%.
These were worst in the {\it U} and $B$ bands, especially for
CCD 1 (the one in the lower left/southwestern corner).  
After determining that the bias levels could not have varied
enough to cause these jumps, we decided to apply a scale factor
correction to each CCD.  While the presence of the galaxy makes it
impossible to assume that the background is flat across the image, we
can still assume that the background is continuous across the chip
gaps.  We made the assumption that CCD 2 is ``truth'' --- this was the
clear choice as most of our standard stars fell on CCD 2.  While the
halo (NE and S) fields required corrections in 
all filters, the CTR field only required corrections in the $U$ and $B$
band.  

We independently checked the validity of these corrections by plotting
the positions of stars in the various color-color diagrams, the
``stellar locus'', for
each chip assuming that the positon and tilt of the locus should not
vary from chip to chip.  Only with these corrections is the stellar
locus seen to be identical in all chips.
After implementing the CCD-dependent corrections, we median combined the
dithered exposures for each field using 3-$\sigma$ rejection to create a
final stacked image. 

\subsection{Photometric Solution}
We observed sixteen standard stars each night from Landolt (1992) 
to perform the {\it UBVRI} photometric calibration.  
Observations in {\it U} and {\it I}\
for the NE and S field were taken
under slightly non-photometric conditions on the first night, but these were
subsequently calibrated on the second night.  The photometric conditions were
similar on nights 2 and 3, allowing us to apply the average photometric
coefficients from both nights to all exposures.
We assumed the mean {\it UBVRI} extinctions for CTIO with values 0.55,
0.25, 0.14, 0.10, and 0.08 magnitudes at the zenith, respectively.

One issue to be wary of when using mosaic cameras is the possibility
that the different CCDs will have significantly different zeropoints and
color terms.  The best way to test this is to measure the same set of
standard stars on each CCD.  This is extremely time-consuming, and was
not practical with the Mosaic's 2.5 minute read-out.  Instead, we used
nine standards on CCD2 to determine the zeropoint in each band, and 
assumed the CCD2 color terms measured by the CTIO staff.  The intra-CCD
zeropoint offsets were determined using the method described in the
previous section.

\subsection{Object Detection and Photometry}
\label{section:detphot}

We chose the widely used
source detection program SExtractor (Bertin 1996) to create catalogs of
objects in our three fields.  SExtractor is well-designed for photometry
of resolved objects on a variable background.  This is the situation we face
when we look for globular clusters superimposed on the smoothly variable galaxy
light.  We chose a mesh size (64 pixels or 16\farcs7) for the 
background estimation algorithm that was large enough 
to not affect the photometry of
unsaturated objects, but small enough to follow the shell structure of
\cena\ as well as subtract the wings of bright, saturated stars.  The
region in the central dust lane was too complex for reliable object
detection and we ignored all objects detected there.

Photometry of all program field objects was done using a $3\arcsec$ 
diameter aperture.  We
corrected these measurements to $14\arcsec$ diameter
apertures using the median growth curve of many bright, isolated stars.
Aperture corrections were measure for each image so that
seeing variations between images would not be an issue.  We also
examined the variation of the aperture correction as a function of
spatial position in the image, and found that it was negligible for a
$3\arcsec$ aperture.  
These corrections were typically on the order of 15\%, or $-0.18$ mag.
The $3\arcsec$ aperture was the best compromise between signal-to-noise,
and the desire for the aperture correction to not vary significantly
across the field-of-view.  SExtractor detected
objects on a summed $V+R$ image, and measured fluxes through the same
apertures in all five bands.

Together in the three fields, we compiled a catalog of $56,674$
unique objects with good photometry---objects not
corrupted by saturation, bleed trails, or edge effects---and $V<23$.

\subsection{Comparison with Published Catalogs}

While there have been many published photometric studies of this region,
one of the most useful comparisons is with the VLT measurements of
probable GCs in Rejkuba (2001).  There are 78 objects for comparison in
$U$ and $V$, 75 of which we detect (the remaining three were likely too
faint).  The $V$ photometry is well-behaved, as
the errors are small for both studies.  The median offsets between the Rejkuba 
and Mosaic photometry are small ($\sim0.03$~mag), although the RMS
scatter ($\sim0.27$~mag) is higher than expected by about a factor of two. 
The $U$ is less consistent because our Mosaic
observations do not go as deep as the VLT data.  There is a
$\sim0.14$~mag offset between the two systems (our magnitudes are
fainter), with $\sim0.3$~mag RMS scatter.  
If significant, this difference could be a result of slightly different
calibrations or aperture corrections.  
Globular clusters at the distance of \cena\ may also be slightly extended in
appearance.  Because our aperture corrections are derived from stellar
profiles, we may underestimate the total flux of the more extended GCs by as
much as $\sim0.1$~mag, although a more typical value would be
$\sim0.06$~mag.  However, this effect is much smaller on 
GC colors since the seeing did not vary much across most of our
observations.  Because the science goals of this study focused mostly on
the relative colors of the GCs, we chose to defer a more precise
treatment of GC light profiles to a later study.

\begin{deluxetable*}{llcclcl}
\tablewidth{0pt}
\tabletypesize{\footnotesize}
\tablecaption{Observing Runs: Spectroscopy \label{table:obsrun_spec}}
\tablehead{
\colhead{Target(s)} & \colhead{Instrument} & 
\colhead{FOV\tablenotemark{a}} & \colhead{Fibers\tablenotemark{b}} &
\colhead{Grating} & \colhead{Res\tablenotemark{c}} & \colhead{Date} \\
\colhead{} & \colhead{} & \colhead{(\arcmin)} & \colhead{} &
\colhead{} & \colhead{(\AA)} & \colhead{}
}
\startdata
PN/GC & AAO/2dF & 120 & 400 & 1200B & 2.2 & 2001 Jan 20 \\
PN/GC & CTIO/Hydra & 40 & 130 & KPGL3 & 4.9 & 2001 Feb 18--20 \\
PN/GC & AAO/2dF & 120 & 400 & 1200V & 2.2 & 2002 Apr 4--6, 9--10 \\
\enddata
\tablenotetext{a}{Field-of-View (diameter)}
\tablenotetext{b}{Number of fibers available for each configuration}
\tablenotetext{c}{Spectral Resolution (\AA\ per resolution element)}
\end{deluxetable*}

Comparisons with the {\it UBVR}
photoelectric photometry of Zickgraf \etal (1990) show that our
photometry is consistent with their published values.
We also checked against the list of confirmed GCs in HGHH92
and other previous work.  This sample 
is considerably brighter than that of Rejkuba (2001) and we detect
all known GCs that are in our field of view.

\section{Building a Catalog of Globular Clusters}
\subsection{A Spectroscopic Survey}

Determining which of our
detected objects are actually GCs in \cena\ is a non-trivial task.
While the total number of GCs in \cena\ is estimated to be 1200--1900
(G.Harris \etal 1984), there
are 17,407 unique objects in our catalog that
fall in the magnitude range that corresponds to the bright half of the
globular cluster luminosity function (GCLF) $14.0<V<20.5$.  Our
faint end cut off for spectroscopy is approximately the magnitude of the
GCLF peak, yet we 
estimate that only 3--$6\%$ of objects in this magnitude range are
GCs.  With such a high level of contamination from foreground stars
(background galaxies are not so important at these relatively bright
magnitudes), it is difficult to know which objects are really GCs.

Some of the
more extended GCs will be resolved in our ground-based imaging
(at the distance of \cena, $1\arcsec\approx17~$pc);
W.Harris \etal (2002) used HST imaging to determine that GCs in
\cena\ have characteristic half-light radii of $0\farcs3-1\farcs0$.  
While size and 
morphology can be used as leverage in selecting high-probability GCs,   
radial velocities are the best way to distinguish between GCs and
foreground stars.  The systemic velocity of \cena\ is 541~\kms\ 
(Hui \etal 1995),
which means that most GCs will have radial velocities larger than those
of Galactic stars.  The central stellar velocity dispersion of \cena\ is
$\sim140$~\kms\ (Wilkinson \etal 1986).  Assuming that the GC system has a
gaussian line-of-sight velocity distribution with a similar dispersion,
introducing a velocity cutoff at
$v_r>250$~\kms\ excludes only the lowest $1.9\%$ of the GC velocity
distribution.  The PN radial velocity distribution is consistent with
this, as only eight of 780 PNe in \cena\ ($1.0\%$) have $v_r < 250$
(Peng, Ford, \& Freeman 2004).  
Even at this velocity cutoff, however, there is a chance that a few
high-velocity Galactic halo stars will
contaminate our sample.  The sun's direction of motion in the disk is
generally away from \cena, so the velocity conversion from the
the Galactic standard of rest to the heliocentric frame
is substantially positive ($+161$~\kms).  Even so, only a small fraction
of the Galactic halo star distribution presents any possible contamination. 

Prior to this work, there were only 77 GCs with published velocities.
These were from a series of papers (VHH81, HHH86, HGHH92), the latter of
which included 32 velocities from the sample of Sharples (1988).  
A number of papers (MAGJM96, HCH99, Rejkuba 2001, W.Harris \etal 2002)
subsequently published lists of probable GC candidates in small
fields based on imaging data alone.  However, it is important to obtain
velocities to be more certain of the nature of these objects.
Modern fiber spectrographs on southern telescopes, such as the 
2-degree Field spectrograph
(2dF) on the 3.9-meter Anglo-Australian Telescope, and the Hydra
spectrograph on the CTIO Blanco 4-meter telescope, are well-suited to
a radial velocity survey in the halo of \cena.  

\subsection{Observations}
\label{sec:gcobs_spec}
We confirmed GCs and refined our target selection over the course of
three spectroscopic observing runs.  Our first and third runs were with
2dF.  The first was in January 2001 during dark time, and the third was
in April 2002 in gray time.  The eponymous two degree
diameter field of view of 2dF is filled with 400 fibers.
Our second observing run was with the CTIO-Hydra spectrograph.  Hydra is
similar to 2dF, except that it has a smaller field of view 
($40\arcmin$ diameter) and $\sim130$ fibers.  
These observing runs are listed in
Table~\ref{table:obsrun_spec}.  In all of these runs, PNe were also targeted.
In addition to our primary aim of discovering new GCs, we tried to 
confirm the nature of published high-probability GCs --- 
the brighter candidates
from MAGJM96, HCH99, and Rejkuba (2001).  We also 
observed many of the HGHH92 GCs for consistency, and in some cases to get
more accurate velocities.

These data were reduced using standard packages and techniques for fiber
spectroscopy.  Details of the reduction are given in a companion paper
on the planetary nebula system of \cena\ (Peng, Ford, \& Freeman 2004).  
Unlike for PNe, we do not gain in
S/N by going to higher dispersion.  For 2dF observations with the 1200V
grating ($2.2$~\AA\ per resolution element), we subsequently smoothed
our GC spectrum by a three pixel boxcar.  The resulting resolution was
comparable to our setup 
with Hydra and did not noticeably compromise our velocity accuracy. 
Sky-subtraction, which is not essential for PNe, was performed on
the GC spectra.

Also unlike in the case of PNe, we performed radial velocity measurements
on GCs using the Fourier cross-correlation method implemented in the
{\tt fxcor} task of the IRAF radial velocity package.  During our
CTIO-Hydra run, we observed the
radial velocity standard HR~5196, which has a published velocity
of $-39.598$~\kms\ (Skuljan, Hearnshaw, \& Cottrell 2000), 
but stars are not always optimal templates for
cross-correlating with globular clusters.  We used HR~5196 to
check the velocity of a high S/N Keck spectrum of an M31 GC, and 
we subsequently used this GC as our velocity template.  

\subsection{Target Selection: Morphology}

\begin{figure}
\centering{\vbox{
\includegraphics[width=3.5in]{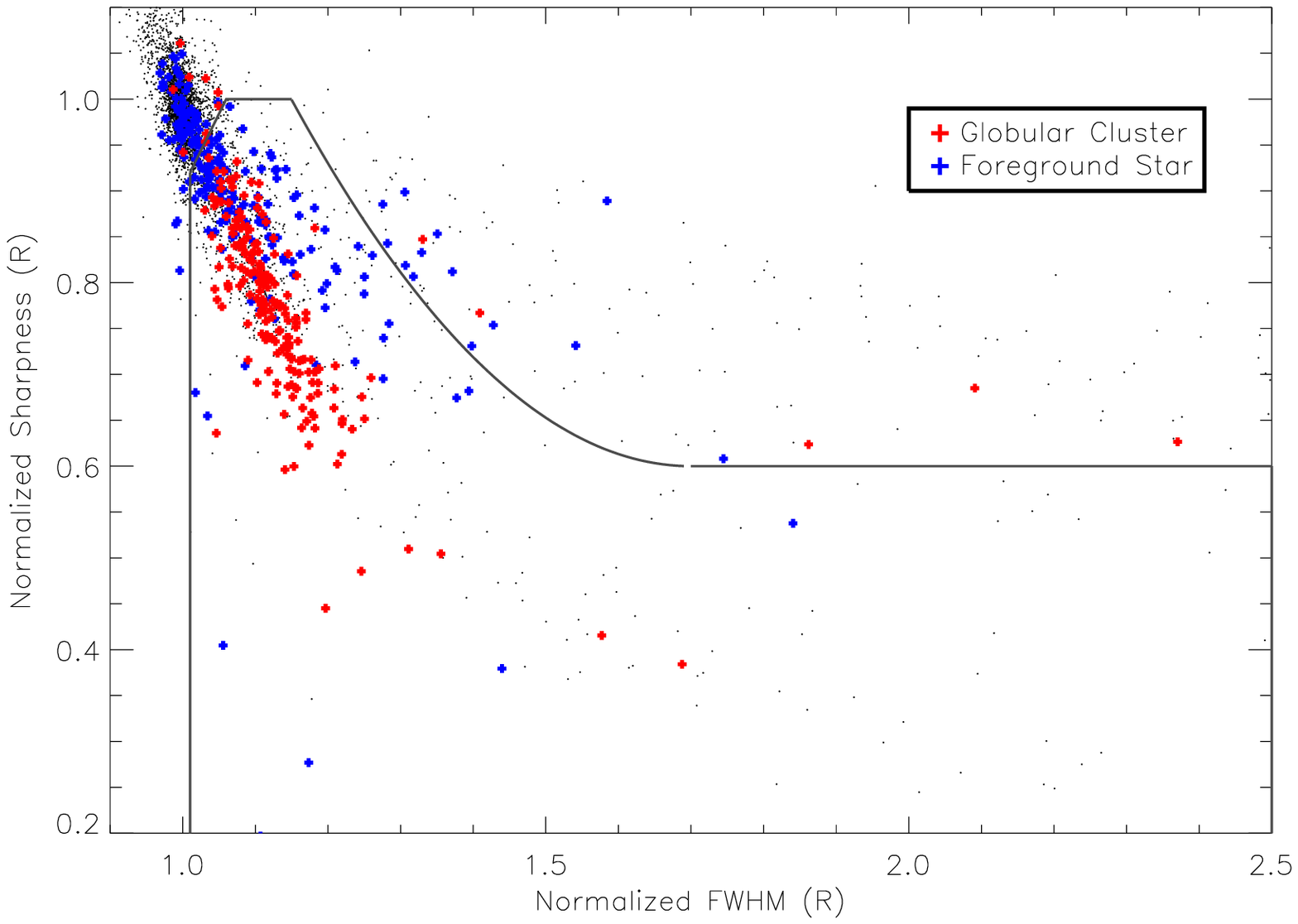}
\includegraphics[width=3.5in]{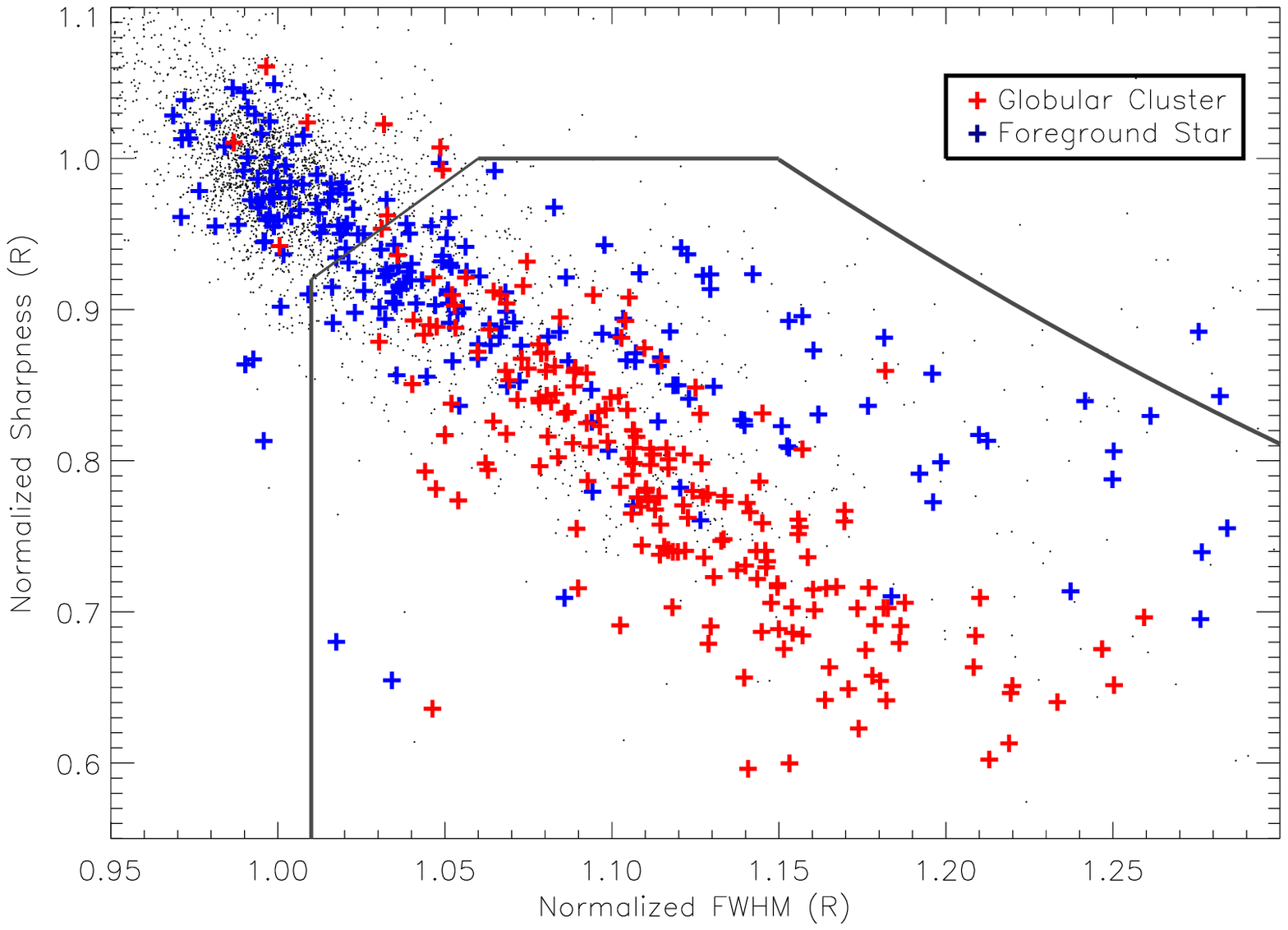}
}}
\caption[GC selection using normalized sharpness and FWHM]
{\small GC selection using normalized sharpness and FWHM.  Sharpness is the
ratio of peak flux to total flux.  Both quantities have been normalized
to local values.  The black line outlines our rather generous selection
boundary.  Objects become more extended as they move to the lower
right.  (a) The top panel shows all candidates for the CTR field as well
as confirmed GCs and stars. (b) The bottom panel shows a zoom in of the
upper left region.
\label{figure:sharpFWHM}
}
\end{figure}

Even with the 400 fibers on 2dF and the 130 fibers on Hydra, we cannot take
a spectrum of every object in the expected magnitude range.  
Thus, it is important that we develop an efficient algorithm to
select likely GCs.
We developed our target selection algorithm using structural parameters
and colors.  With the best seeing in each field being nearly $1\arcsec$,
it is possible to resolve some of the more extended GCs in our imaging.
We use two structural parameters, sharpness and full-width at half
maximum (FWHM), in conjunction
with ellipticity to identify round, extended objects in our catalog.  We
define sharpness to be the ratio of flux in the brightest pixel to the
``total'' flux --- that measured within the SExtractor automatic
Kron aperture.  This is essentially a difference of aperture magnitudes
that measures the concentration of the object.  This approach benefits
from a nicely oversampled PSF (5 pixels per FWHM).
Because the point spread function varies across the
field-of-view, we normalized both sharpness and FWHM to the local
value.  Thus, objects that are extended
have normalized sharpnesses less than unity, and normalized FWHMs greater
than unity.

Figure~\ref{figure:sharpFWHM} shows plots of normalized sharpness versus
normalized FWHM for the CTR field.  The bottom plot is a zoomed-in
version of the top.  The black dots are objects in the
CTR field with $V<20.5$ and $U<23$.  Most of the objects cluster around
(1,1), as we would expect for a sample dominated by stars.  However,
there is a fan of points toward lower sharpness and higher FWHM.  
Overplotted in red are {\it all} spectroscopically confirmed GCs, 
and in blue are all
spectroscopically confirmed foreground stars.  This is not the training
set we had to work with initially, but we plot the end-product catalog
for the purpose of demonstration.
It is also important to
remember that when we first decided on our cuts, we
based our selection on the 63 HGHH92 GCs and some of the
brighter high-probability GCs from Rejkuba (2001).  We did not have a
sample of confirmed 
foreground stars until after our first spectroscopic run.
Results from each observing run were used to improve the selection
for subsequent runs.  

\begin{deluxetable*}{ll|c|ccccc|c}
\tabletypesize{\scriptsize}
\tablecaption{Spectroscopic Follow-up of GC candidates \label{table:gcresults}}
\tablewidth{0pt}
\tablehead{
\colhead{} & \colhead{} & \colhead{Total} &
\multicolumn{5}{c}{Newly Confirmed GCs\tablenotemark{b}} &
\colhead{Known GCs\tablenotemark{c}} \\
\colhead{Facility} & \colhead{Dates} & \colhead{Targets\tablenotemark{a}} &
\colhead{Mos} & \colhead{HHH} & \colhead{R01} & \colhead{MAGJM} & 
\colhead{HCH} & \colhead{HGHH}
}
\startdata
AAO/2dF & 2001 Jan 20 & 120 & 2 & 0 & 0 & 0 & 0 & 28 \\
CTIO/Hydra & 2001 Feb 18--20 & 254 & 88 & 2 & 28 & 4 & 6 & 7 \\
AAO/2dF & 2002 Apr 4--6, 9--10 & 59 & 19 & 0 & 0 & 0 & 0 & 8 \\
\enddata
\tablenotetext{a}{The total number of GC candidates plus known GCs
targeted in this observing run.  Equal to the sum of numbers in
subsequent columns plus number of foreground stars observed.}
\tablenotetext{b}{Number of new GCs that resulted
from observations of previously unconfirmed candidates originating from
each of the four possible sources: Mosaic (our photometry), HHH86, Rejkuba
(2001), MAGJM96, and HCH99.}
\tablenotetext{c}{Number of previously confirmed GCs that we re-observed.}
\end{deluxetable*}

The GCs are clearly in a locus offset from the stars.  This
morphological selection bias has been built into the
sample---most known GCs in earlier work and our study were
selected at least in part because of their spatial extent---yet it shows
that at least some GCs can be separated from stars given the seeing in
our images.  Not all stars are concentrated at (1,1), as there is a tail
of stars to higher FWHM.  These stars, however, are typically sharper
for their FWHM than are the GCs.  Upon visual inspection, we found that 
the locus of sharp but extended points primarily consist of unresolved
blends.  One disadvantage of using SExtractor is that with only one
iteration of object detection and no PSF subtraction, it cannot detect faint
stars that are very close to brighter ones.  The bottom plot illustrates
more clearly how the loci of stars and GCs separate in sharpness at
larger FWHM.

The lines in Figure~\ref{figure:sharpFWHM} denote the area within which
we accepted an object as a GC candidate for our second and third
spectroscopic runs
(there was no sharpness/FWHM cut for the initial run).  
First, we exclude most of the
stellar locus.  The top line that asymptotes
to a normalized sharpness of 0.6 excludes the most
egregious blends from being targeted (although it may also exclude
star-GC blends).  While the GCs actually form a fairly tight sequence in
this plot, we chose to be conservative and allow more stars into the
selection area to increase our completeness.  
The last morphological cut was a
relatively loose one that rejected all objects with ellipticities
greater than 0.4.  Almost all GCs have ellipticities less than 0.1, so
this cut simply rejected obvious background galaxies, and also served to
eliminate partially resolved pairs of objects.

\subsection{Target Selection: Optical Colors}

In addition to selecting on size and shape, we instituted color cuts to
increase our efficiency.  Globular clusters are old, single-age stellar
populations that have optical flux predominantly produced by stars from the
main sequence turnoff to the red giant branch.  The integrated
colors of these stellar populations are produced by combinations of stars
with types F through K (and maybe also A-type giants for GCs with blue
horizontal branches).  The result is that the spectral energy
distribution of a GC is broader than that for any individual
star.  Hence, with a long enough wavelength baseline and sufficiently
accurate photometry, we should be able to distinguish GCs from stars
using colors alone.

\begin{figure}
\centering{\vbox{
\includegraphics[width=3.5in]{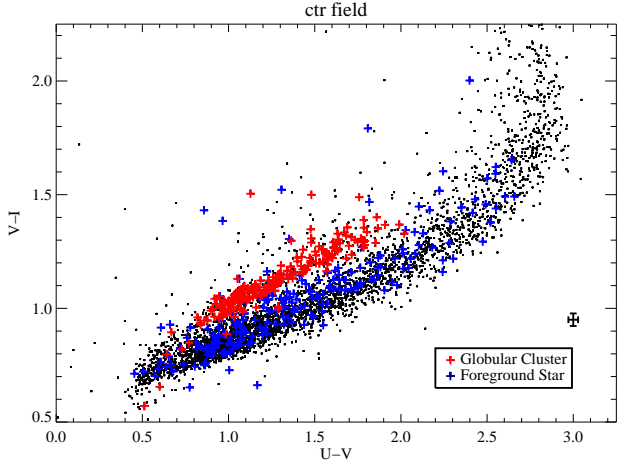}
}}
\caption[Color separation for GCs and stars in $UVI$]
{\small Color separation for GCs and stars in $UVI$ 
Black dots are objects in our CTR
field with $14<V<20.5$.  The error bars are representative for bright
object photometry.
Globular clusters are offset from the stellar locus in the way one would
expect if they had a composite stellar population.  Discrepant GCs with
(\vi)$\sim1.5$ have large photometric errors.
\label{figure:colorselect}}
\end{figure}

Figure~\ref{figure:colorselect} shows a (\uv)--(\vi) color-color
diagram---the combination of colors that offers the widest
wavelength baseline for our data.  The black dots are objects in our CTR
field with $14<V<20.5$.  The dense concentration of points outline the
$UVI$ stellar locus.  As in Figure~\ref{figure:sharpFWHM}, all confirmed stars 
and GCs are overplotted as blue
and red plus symbols, respectively.  Although the division is not
perfect, there is a clear separation between GCs and stars in $UVI$ color
space---at a given \uv, GCs are redder in \vi.  
$U$-band photometry, though expensive, is valuable for the separation
of GCs from stars.

\subsection{Spectroscopic Yields}

The results of these three spectroscopic runs are summarized in
Table~\ref{table:gcresults}.  The total numbers of clusters listed will
be larger than the numbers in the final catalog because some were
observed in multiple runs.
We took slightly different targeting strategies in each run.  For the
first 2dF observations, GC candidates were piggybacked onto observations
of PNe.  PNe had the highest priority while GC candidates were used to
fill empty fibers in the uncrowded halo regions.  Most of these
candidates were in the NE field, and we used only broad color cuts,
no morphology cuts, and a magnitude limit of $V<20$.  This selection
produced very few new GCs (only a 2\% yield), because the color cut
included a large portion of the stellar locus, and the candidates were
restricted being the NE halo field where the density of GCs is low.
Our three night Hydra run also combined PNe and GCs, although GCs
received the higher priority.  This time, we implemented a cut on
normalized sharpness and FWHM, combined with a weak cut on color.
Our third observing run was 
another where we targeted both PNe and GCs.  For this fiber
configuration, we implemented a combined $UVI$ and morphology cut, and
also extended the magnitude ranges to $V<20.5$.  The yield of new GCs
was $\sim40$\% for both of these runs, showing that careful selection
criteria can increase efficiency by an order of magnitude.
The results of our observing program are shown in the selection diagrams of
Figures~\ref{figure:sharpFWHM} and \ref{figure:colorselect}.

\section{Galaxy Background Subtraction}

For the purposes of measuring line indices (and to a lesser extent,
radial velocities), it is important to determine the fraction of
light that went through the fiber that is from the
targeted globular cluster.  The nature of
fiber spectra is such that background subtraction is not local.  
Typically, thirty or more sky fibers are placed at positions free of objects,
and their spectra are combined to create a master sky spectrum.
This spectrum is then scaled to each individual object spectrum, either
using sky lines or measured throughput differences, and then
subtracted.  This approach works well (to within 1--2\%) when the
background is relatively constant.  However, in the case of our
observations around \cena, the background in regions near the galaxy
center is a spatially varying combination of the sky and the
unresolved galaxy light.  This is not a problem for slit spectra where
the local background is determined for each object individually in
spatially adjacent regions.  This does become a problem for fiber
spectra where no single composite sky spectrum is representative for the
entire field of view.  Since sky fibers are given lower priority and
are often placed away from the galaxy center (where there is less
crowding from program objects), sky fibers typically contain little
galaxy light.  Thus, the spectra of faint, centrally located GCs may
have a significant contribution from the field light of the galaxy.
For line index measurements in particular, it is important that these
GCs have the galaxy contribution removed from their spectra.

\begin{figure}
\plotone{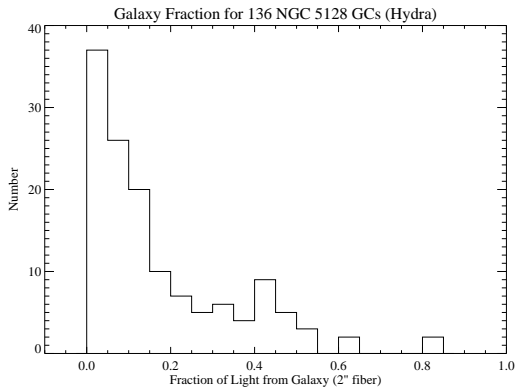}
\caption[Fraction of Galaxy Light Down Fiber for GC spectra (Hydra sample)]
{\small Fraction of galaxy light down each fiber for GC spectra (Hydra sample).
Using $B$ and $V$ images, we calculated the fraction of light that
went down each 2\arcsec\ diameter fiber that originated from the galaxy's
unresolved stellar light.  Most spectra are relatively free of galaxy
contamination. 
\label{figure:gfrac_hist}}
\end{figure}

With the GC fiber spectra alone, this would be a difficult problem.  The
many different aspects of our data set, however, permitted us to develop
a workable solution.  First, when creating the composite sky spectrum,
we used only those sky spectra that showed no visible contribution from
the galaxy so that it was pure sky.  This was not a problem as, again,
most sky fibers were placed well away from the bright parts of the
galaxy.  After sky subtraction, the reduced program spectra are some
combination of cluster and galaxy.  Using our imaging in the $B$ and $V$
bands, we measured the fraction of 
light down each fiber (2\arcsec\ diameter) that should belong to the
galaxy.  The distribution of this galaxy fraction for the Hydra-observed
GCs is shown in
Figure~\ref{figure:gfrac_hist}.  It is reassuring that most of the GCs
have only a low level of galaxy contamination (median of 11\%), 
and almost all have contamination lower than 50\%.

In order to subtract the galaxy's contribution to each spectrum, we
created a template galaxy spectrum.  By summing the sky 
fibers that have visible galaxy signal even when the master sky spectrum
has been subtracted---those targeting ``sky'' regions toward the center
of the galaxy---we obtained a high S/N galaxy composite.
By scaling this composite to the galaxy fraction computed with the
imaging data, we are able to remove the underlying galaxy contribution.

The last variable is the velocity and velocity dispersion of the galaxy
at the location of each fiber.  Because the rotation in the mean
velocity field can cause the galaxy spectrum to shift by $\pm1.5$\AA, it 
is important that the scaled spectrum be redshifted to the proper
velocity before subtracting it.  For this, we use the mean velocity of
the planetary nebulae (PNe) at each fiber location (Peng, Ford, \& Freeman
2004).  Because PNe trace the kinematics of the old stellar population,
and the field should be smoothly varying, this is a valid method of
estimating the local mean velocity field.  The velocity dispersion can
be assumed  to be constant for these purposes.

\begin{figure}
\plotone{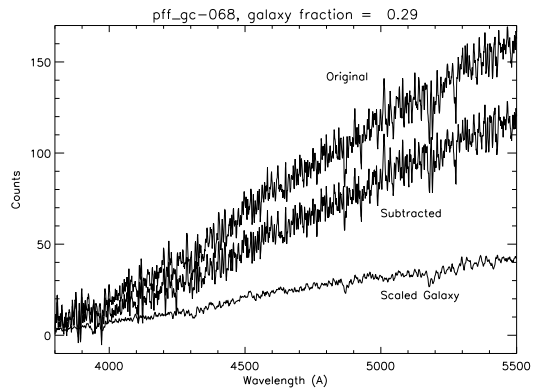}
\caption[Galaxy subtraction example]
{\small An example of galaxy subtraction.  This star cluster's spectrum
has a high contamination from galaxy light.  The original spectrum is
shown in black (top) with the scaled and redshifted galaxy spectrum in
red (bottom), and the final subtracted spectrum in blue (middle).
Notice how the galaxy template is broader and slighted shifted from the
cluster spectrum.
\label{figure:pff_gc-068}}
\end{figure}

After redshifting and scaling the composite galaxy spectrum for each GC,
we removed the bulk of the galaxy's signature in each spectrum.
While this technique removes zeroth order contributions from the galaxy,
it can introduce small errors.  If the seeing is different
during the spectroscopic observing run than during the imaging, then the
computed galaxy fraction will be slightly off.  For example, if the
seeing for the imaging was $1\farcs2$ but the seeing for spectroscopy
was $1\farcs5$, then a GC with a calculated galaxy contamination of 11\%
will instead have a true contamination of 13\% due to more of the GC
light being outside the fiber aperture.  Also, since the composite
galaxy spectrum is a mixture of light from different regions in the
galaxy, it does not take into account spatial variations in
metallicity and age that may cause the background to vary.  However,
without a spectroscopic measurement of the local background, by either
using slits or chopping to local sky, this is probably the best
background subtraction we can achieve.  
Figure~\ref{figure:pff_gc-068} shows an example of this subtraction in 
the $V=18.0$ cluster, pff\_gc-068, a
case where the galaxy fraction is fairly high ($\sim0.3$).   
In addition, the galaxy fraction distribution
(Figure~\ref{figure:gfrac_hist}) shows that few GCs are like the
example in Figure~\ref{figure:pff_gc-068} and so this procedure is
important for only a subset of GCs.  

Over half of our spectroscopic sample with Mosaic photometry (136 GCs) 
was observed with CTIO/Hydra.  This is the only observing
run for which we obtained high signal-to-noise spectra of the galaxy
and Lick/IDS standard stars.  These spectra also had the largest
wavelength coverage (3800-5500\AA).  Therefore, we only performed galaxy
subtraction and line index measurements on the CTIO/Hydra sample.  
All Hydra spectra had their radial velocities re-measured after galaxy
subtraction, but the changes were minimal --- a few \kms\ at most.

\vspace{1cm}

\section{Catalog}
\subsection{The Globular Cluster Catalog}
\label{sec:gccat}

The final globular cluster catalog contains 215 unique globular clusters
in \cena\ from various sources.  Of these, 138 are newly confirmed GCs
that have no velocities in the published literature.
In each of our spectroscopic
runs, we re-observed objects from previous runs to provide a consistent
velocity zeropoint.  We also re-observed GCs with published velocities
(HGHH92) for comparison.  These repeat
observations are important for creating a consistent velocity catalog,
and for quantifying our velocity errors.  We present these data in
Table~\ref{table:gcrepeat}. 

\begin{deluxetable}{lcccc}
\tablecaption{GC Velocity Offsets and Errors from Repeat Observations 
\label{table:gcrepeat}}
\tablewidth{0pt}
\tablehead{
\colhead{Velocity Source} & \colhead{Offset} & 
\colhead{RMS} & \colhead{No.\ Overlap} & \colhead{$<V>$\tablenotemark{a}} \\
\colhead{} & \colhead{(\kms)} & \colhead{(\kms)} & \colhead{} & \colhead{} 
}
\startdata
Hydra 2001 & 0 & 0 & N/A & N/A \\
2dF 2001 & $2\pm15$ & 48 & 10 & 19.1 \\
2dF 2002 & $-14\pm10$ & 65 & 22 & 19.4 \\
HGHH92\tablenotemark{b} & $-26\pm24$ & 85 & 12 & 18.6 \\
HGHH92\tablenotemark{c} & $-26\pm14$ & 73 & 28 & 18.7 \\
\enddata
\tablenotetext{b}{Mean $V$ magnitude of overlapping objects}
\tablenotetext{b}{Comparison with 2dF 2001}
\tablenotetext{c}{Comparison with Hydra}
\end{deluxetable}

We chose to compare all velocities to the Hydra observations, the only
observing run for which we obtained a
radial velocity standard.  The offsets between Hydra and the two 2dF
runs are negligible, and consistent with zero.  The offset between our
data and the published velocities in HGHH92 is approximately $-26$~\kms.
While the dispersion for these HGHH92 GCs is considerably higher, the
offset is marginally significant.
The dispersions for the repeat velocity data vary from 48--85~\kms.
This can be explained by the data quality of the different repeat samples.
The mean $V$-magnitude of the twelve overlap objects between the Hydra and
2001 2dF runs is 0.6 mag brighter than the mean $V$-magnitude of the
overlap objects between Hydra and the 2002 2dF run.  
Many of the 2002 2dF observations were also taken during
fairly bright moon.  Assuming that the Hydra and 2dF observations have
similar errors that add in quadrature, we derive a velocity error of
34~\kms\ for objects with $V<19$.  This is consistent with the median
velocity error of our sample as derived from the Fourier cross
correlation technique.  For fainter GCs, however, the velocity
error can be as large as 70~\kms.

The large velocity dispersion between our data and the HGHH92 published
velocities are not likely due to low S/N.  The mean magnitudes of these
overlap GCs are well in the high S/N regime.  Given the accuracy of our
other repeat measurements, it is likely that
the intrinsic errors of the HGHH92 velocities are $\sim65$~\kms.
Therefore, when incorporating the HGHH92 GCs into our final catalog, we
always replace the published velocity with our own measurement if it
exists.  If we did not observe a given HGHH92 GC, then we use the
published velocity with an offset of $-26$~\kms, as shown in
Table~\ref{table:gcrepeat}. 

We visually evaluated the quality of each spectrum and radial
velocity.  For objects with multiple observations, we retained
the velocity derived
from the highest S/N spectrum.  In cases where the S/N was
comparable for all observations, we used the weighted 
mean of the observed values.

In Table~\ref{table:gcv}, we present the final GC velocity catalog.
For each object, we list its coordinates, \ubvri\ photometry (if it
exists) and its final heliocentric radial velocity.  Column 1 is the
name of the object; Columns 2 and 3 are the RA and Dec coordinates in
J2000.0 on the USNO-A2.0 astrometric reference frame; Column 4 is
$R_p$, the projected radius in arc minutes; Columns 5 and 6 are
$X$ and $Y$, \cena-centric coordinates that are arc minutes along the
photometric major and minor axes, respectively, where the major axis is
taken to have a position angle of 35\degr; Column 7 is the $V$
magnitude; Columns 8--11 are the \ub, \bv, \vr, and \vi\ colors; Columns
12--16 are the associated photometric errors; Column 17 is .  
$V_{h}$, the heliocentric velocity; Column 18 is $V_{h,err}$, the error
in the measured velocity, or if the velocity is from the literature,
then we assume an error of 65~\kms; Column 19 is $E_{B-V}$,
the reddening for each GC as derived from the maps of Schlegel,
Finkbeiner, \& Davis (1998).  Column 20 is the 
``Run'', which designates during which observing run(s), if any, the
object was observed, where the 
runs are numbered such that ``1'' is the 2dF 2001 run, ``2'' is the
Hydra 2001 run, and ``3'' is the 2dF 2002 run.

We always give GCs the primary designation from the
earliest publication in which it appears (except for GP80-1), and we
have done our best to match duplicates.  Rejkuba (2001) GCs are labeled
by and ``R'', then by a 1 or 2 depending on whether they were in her
fields 1 or 2, then by the two digit number that she gave them.
All GCs that have a VHH81 or HGHH designation had previously published
velocities.  All GCs with an HHH86 designation, except for HHH86-13 and
HHH86-15, also had published velocities.  The remaining GCs are newly
confirmed.  A few GCs have multiple names in the published literature.
This is true for a few Rejkuba (2001) GC candidates, where R226 is
HHH86-15, R123 is HHH86-38, R281 is HGHH-12.
The GC from Graham \& Phillips (1980) is also HGHH-07.  

\begin{figure*}
\plotone{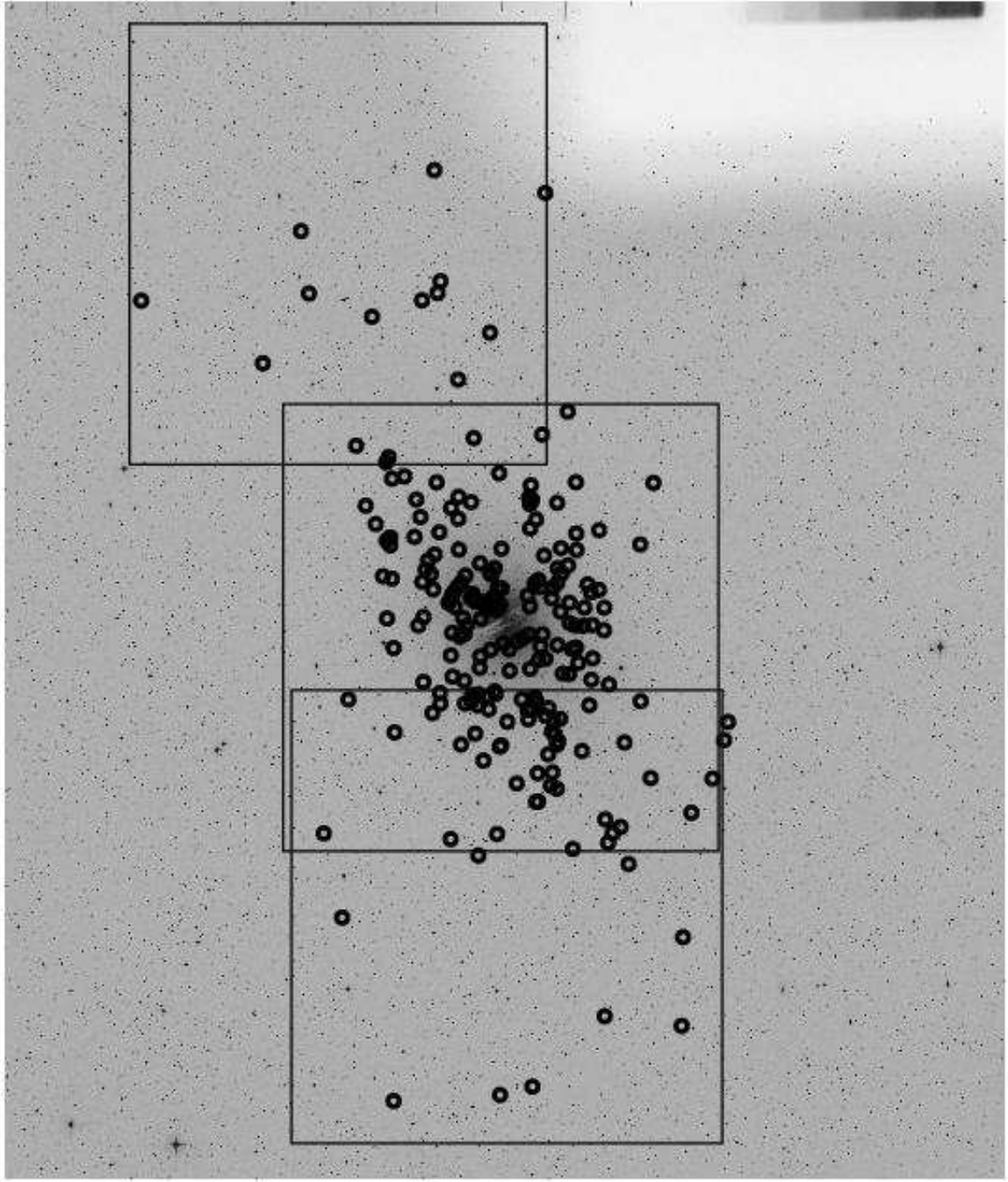}
\caption[GC Positions and Mosaic Fields over DSS image]
{\small GC positions and Mosaic fields overplotted on a DSS image of \cena.
\label{figure:gcspatial}}
\end{figure*}

Figure~\ref{figure:gcspatial} shows the locations of all confirmed GCs
with respect to the galaxy and the three Mosaic fields.  While the
farthest GCs are at projected radii greater than 40~kpc, slightly over
half are contained within $2r_e$, where $r_e = 5.2$~kpc.  The median
projected radius of the GCs in our sample is $\sim9$~kpc.

\subsection{Foreground and Background Objects}

\begin{figure}
\plotone{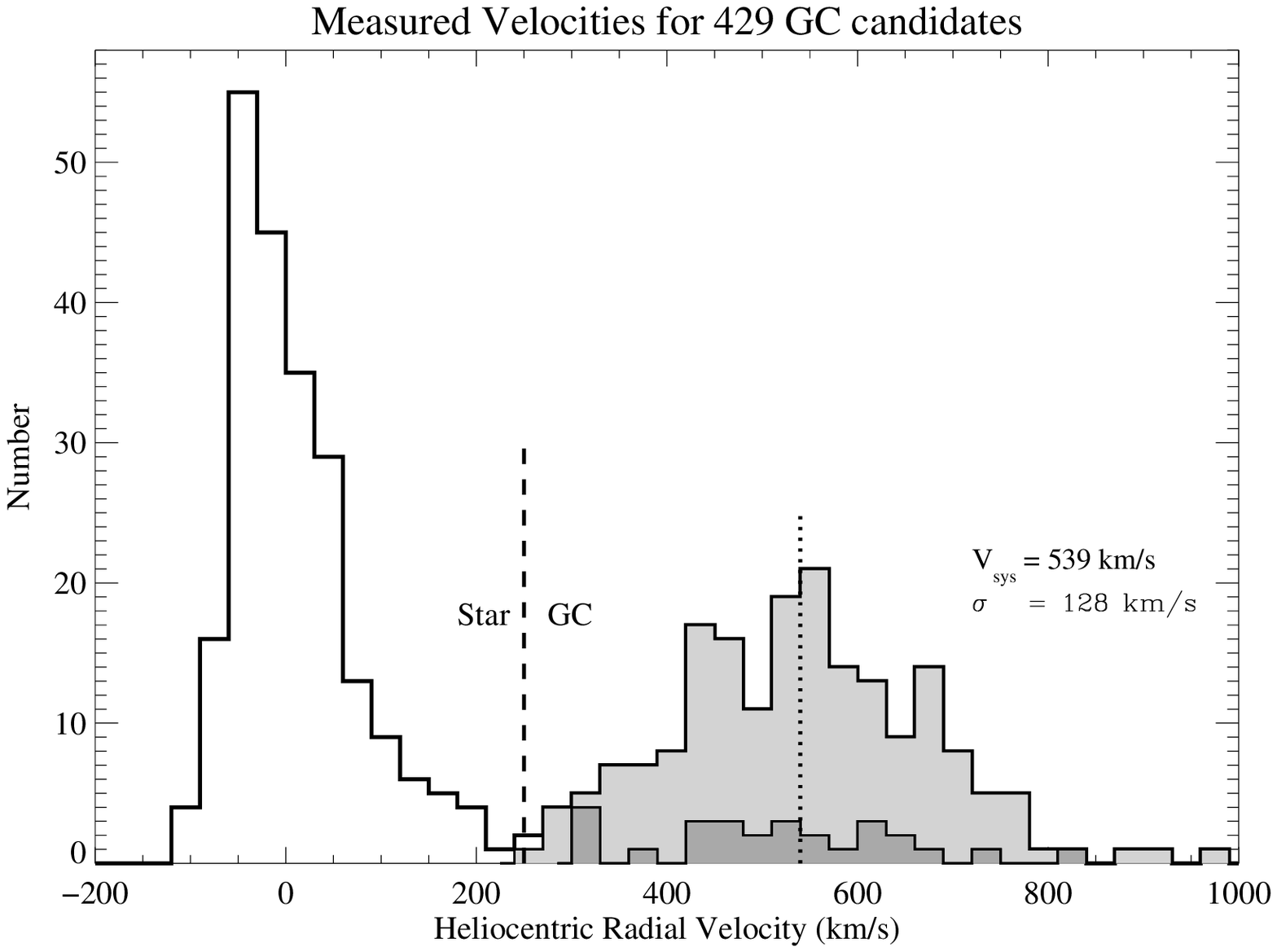}
\caption[Velocity Distribution of Cluster Candidates]
{\small Velocity Distribution of Cluster Candidates.
A histogram of velocities for all objects
targeted as globular cluster candidates.  The foreground star
contamination is apparent, as is the population of true Cen~A globular
clusters.  While there may be a few misclassified objects, the two
populations are well-separated in velocity.
Objects at radial velocities between 250 and 1000 km/s
are classified as clusters.  The black line is the histogram for objects
observed in our program, with the light gray shading representing the
GCs.  The underlying dark gray histogram is the fraction of our observed 
sample that was previously confirmed (from VHH81, HHH86, and HGHH92).  
This does not represent the full sample because we did not reobserve all
known GCs.  The mean systemic velocity of the GC system is 539~\kms,
which is observationally indistinguishable from the systemic velocity
derived from the PNe. 
\label{figure:gchists_rv}}
\end{figure}

Table~\ref{table:fgs} lists all confirmed foreground stars in the halo 
of \cena.  This table is included so that future
astronomers looking for GCs in \cena\ need not experience the
disappointment of re-observing them.  Note that while almost all observed
Rejkuba (2001) candidates are true GCs, f1.gc-19 (R119) is actually a star.
The column labels are the same as for Table~\ref{table:gcv}, except that
the columns containing $R_p$, $X$, and $Y$ are not included. 
The combined velocity distributions of these two samples is
shown in Figure~\ref{figure:gchists_rv}.

We also list GC candidates that were neither GCs nor
stars.  This was the case for very few objects, but in some cases,
candidates turned out to be background galaxies.  These are listed in
Table~\ref{table:galx}.  Redshifts were determined from one or more
spectral features such as [\ion{O}{2}], [\ion{O}{3}], or Ca H and K, and
are good to within 0.01.  Column labels are the same as for
Table~\ref{table:fgs}.

\subsection{Matches with X-Ray Point Sources}

Using Chandra data, Kraft \etal (2001) produced a list of 246 X-ray
point sources within a couple effective radii in \cena.  Many X-ray
point sources in old stellar populations have optical
matches that are globular clusters (Kundu, Maccarone, \& Zepf 2002),
and it is believed that GCs may be the dominant sites for the formation
of low-mass X-ray binaries.  We present 52 optical
matches to the Kraft \etal\ catalog from our photometry.  We matched objects
within a $3\arcsec$ radius, that had $17<V<21.5$ and were not in the
dust lane.  Thirteen of these are
known GCs, and the others are likely to also be GCs.  The full list is
presented in Table~\ref{table:xraymatch}. 
Column labels are the same as for Table~\ref{table:fgs} except that
Column 1 contains the ID from Kraft \etal (2003), Column 2 is the ID
given from any previous GC survey, and we do not include columns for
$R_p$, $X$, $Y$, $V_h$, or Run.  Like others, we find that 
the optical
counterparts to X-ray point sources are more likely to have colors
consistent with metal-rich GCs than metal-poor GCs.
A thorough analysis of the nature of the optical-X-ray sources in \cena\
is presented with complementary data by Minniti \etal (2003).

\subsection{Serendipity: White Dwarfs and QSOs}

In addition to looking for old GCs, we used some fibers to specifically
target blue objects in the \cena\ field in the hopes of finding
young clusters in the halo, like the one in the young tidal stream
described in Peng, Ford, \& Freeman (2002).  We
implemented a simple magnitude cut $V < 20.0$, and color
cut $(B-V)_0 < 0.285$, assuming a foreground reddening 
$E(B-V) = 0.115$~mag (Schlegel \etal 1998).  While we
did not find any unambiguous young clusters in the halo, we did discover
low-redshift emission-line galaxies, metal-poor dwarf stars, white
dwarfs and QSOs.  Three white dwarfs and seven QSOs are listed in
Table~\ref{table:blueobj}.  Columns are similar to
Tables~\ref{table:gcv}--\ref{table:xraymatch}, except the redshift of
QSOs is listed in Column 9.  The white dwarfs were identified by their
broad \Hb\ absorption, and the QSOs by their \ion{Mg}{2} emission.  
Occasionally, intervening \ion{Mg}{2} absorbers were visible blueward of
the \ion{Mg}{2} emission.  None of these lines are likely to be
Lyman-$\alpha$, because
in all cases, there is detectable continuum blueward of the emission line.
Redshifts were determined using the \ion{Mg}{2} emission line, and are
good to 0.05.  Our relatively narrow wavelength
range restricted us to identifying QSOs with redshifts around
$z\sim0.7$.  Although these QSOs are not especially bright ($V>19$),
they seem to have smooth continua and may eventually be useful for
probing the interstellar medium in the halo of \cena.

\vspace{1cm}

\section{The Reported Planetary Nebula in HGHH-G169}

\begin{figure}[t]
\plotone{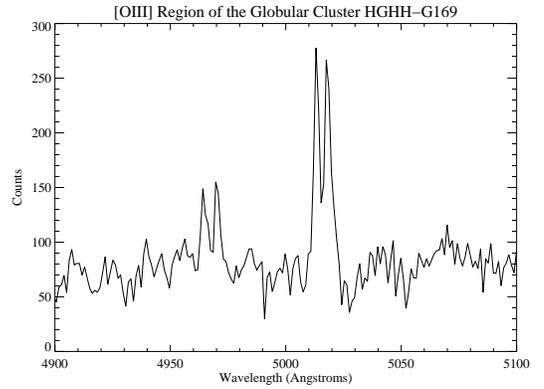}
\caption[2dF spectrum of HGHH-G169, $\lambda5007$\AA\ region]
{\small A 2dF spectrum of HGHH-G169.  We zoom on the
[OIII]$\lambda\lambda4959,5007$ lines to show that they are split by
roughly 5\AA\ (300~\kms). \label{figure:g169}}
\end{figure}

In the course of obtaining spectroscopy for GCs in \cena, 
Minniti \& Rejkuba (2002, MR02) found that the GC HGHH-G169 had strong
[\ion{O}{3}] emission in its spectrum.  They concluded that it was a
planetary nebula within the GC---the first of its kind in an elliptical.  
PNe are extremely rare in Galactic globular
clusters, and it would be exciting if more GC-associated PNe
could be located in other galaxies.

During the course of our survey, we also observed HGHH-G169.  Like MR02,
we targeted it (in our 2001 2dF run) because it is listed in HGHH92.  
We also noticed strong [\ion{O}{3}] emission in the GC spectrum, and
thought it might be a PN.  However, we were somewhat puzzled because both
[\ion{O}{3}] lines ($\lambda\lambda4959,5007$) are double-peaked, with the 
separation between the peaks being about 5\AA\ (300~\kms).  The [\ion{O}{3}]
region of the spectrum is shown in Figure~\ref{figure:g169}.

We carefully checked the raw data to test the validity of this split.  The
split lines are seen in all five of the pre-combined frames, and the arc
calibrations show only single peaks.  Hence, we are convinced that this
split is real.  Moreover, it is unlikely that MR02 would have been able 
to resolve these lines.  With the Boller and Chivens spectrograph on the
Magellan Baade telescope, they likely had 6\AA\ instrumental resolution.
Our 2dF resolution was 2.2\AA. 

If we assume that the line splitting is due to seeing the front and back
sides of an expanding shell (with an expansion velocity of ~150 km/s),
then it is unlikely that this object is a PNe. Planetary nebulae
normally have expansion velocities of only a few tens of \kms.  
In addition, the spectrum displayed in MR02's
Figure~2~(top) shows hints of \ion{He}{1}$\lambda5876$ and
[\ion{O}{1}]$\lambda6300$ emission.  If these low-excitation
lines are confirmed, then this object is unlikely to be a PN.  We think
that the different ionization states and the fast expansion velocity makes
it possible that this is a supernova remnant, which would also be very
interesting.

We also have ground-based [\ion{O}{3}] narrow-band and continuum images of this
region from the survey of H93.  Despite its strong [\ion{O}{3}] flux, 
this object would not have met H93's or our selection criteria because
we require that there be no detection in the continuum image; in this
case, there is a bright GC visible in the off-band at that position.
Upon close inspection, the source of the [\ion{O}{3}]
emission is offset by $\sim1\farcs4$ northeast of the center of 
HGHH-G169, which translates to a physical distance of 24~pc.  While this
is still close enough to be associated with HGHH-G169, the object would
need to be toward the outskirts of the cluster.

Whether or not this object is a PN, it is still interesting and highlights
the need to inspect GC spectra for other features.  While some of the
GCs near the dust lane have spectra that contain low excitation emission
lines, this is likely to be due to the star formation that is ongoing in
the region rather than from emission within the GC.  None of our
other spectra showed signs of having a planetary nebula.  As MR02
discuss, these numbers are starting to place some statistically
interesting constraints on the formation of PNe in globular clusters.

\section{Summary}

We conducted an optical imaging and spectroscopic survey for globular
clusters across $\sim 1\degr$ of sky around \cena.  Using \ubvri\
photometry, size, and morphological information, we developed an efficient
algorithm for selecting likely GCs candidates for spectroscopic
follow-up---a necessity given the large number of foreground stars.  We
obtained radial velocities for over 400 objects.  Of these, we identified
102 previously unknown GCs, confirmed the nature of 24 GCs from Rejkuba
(2001), 6 GCs from HCH99, 4 GCs from MAGJM96, and 2 from HHH86, 
providing 138 new GC
velocities.  We also obtained new 
velocities for 25 previously confirmed GCs from HGHH92 and HHH86.  The
total number of confirmed GCs in \cena\ is now 215.

We present a spectrum of HGHH-G169, which shows an interesting split
emission line feature.  It is unlikely to be a planetary nebula, as MR02
suggest, but it could be a supernova remnant, and certainly merits a
deeper spectrum.

\acknowledgments

E.\ W.\ P.\ acknowledges support from NSF grant AST 00-98566. 
H.\ C.\ F.\ acknowledges support from NASA contract NAS 5-32865
and NASA grant NAG 5-7697. We thank the staffs at CTIO and the AAO for
their invaluable help during our observing runs.  E.\ W.\ P.\
thanks Alan Uomoto and Christy Tremonti for useful discussions.
This research has made use of the NASA/IPAC Extragalactic
Database (NED), which is operated by the Jet Propulsion Laboratory,
California Institute of Technology, under contract with NASA. 
This research has made use of the SIMBAD database, operated at CDS,
Strasbourg, France.

\begin{turnpage}
{\centering
\begin{deluxetable}{lccrrrcccccccccccrcc}
\tabletypesize{\tiny}
\tablecaption{Catalog of \cena\ GCs with Radial Velocities.}
\tablewidth{0pt}
\tablehead{
\colhead{Name} & \colhead{RA} & \colhead{Dec} &
\colhead{$R_{p}$} & \colhead{$X$} & \colhead{$Y$} &
\colhead{$V$} & \colhead{$U$-$B$} & \colhead{$B$-$V$} &
\colhead{$V$-$R$} & \colhead{$V$-$I$} &
\colhead{$\sigma_U$} & \colhead{$\sigma_B$} & 
\colhead{$\sigma_V$} &
\colhead{$\sigma_R$} &\colhead{$\sigma_I$} &
\colhead{$V_{h}$} &
\colhead{$V_{h,err}$} &
\colhead{$E_{B-V}$} &
\colhead{Run} \\
\colhead{} & \colhead{(J2000)} & \colhead{(J2000)} &
\colhead{(\arcmin)} & \colhead{(\arcmin)} & \colhead{(\arcmin)} &
\colhead{mag} & \colhead{mag} & \colhead{mag} &
\colhead{mag} & \colhead{mag} &
\colhead{mag} & \colhead{mag} & 
\colhead{mag} &
\colhead{mag} &\colhead{mag} &
\colhead{km/s} &
\colhead{km/s} &
\colhead{$E_{B-V}$} &
\colhead{} 
}
\startdata
pff\_gc-001 & 13:23:49.62 & $-$43:14:32.0 &  22.38 & $ -21.290$ & $   6.903$ & 18.91 &  0.47 &  1.01 &  0.55 &  1.31 & 0.09 & 0.02 & 0.01 & 0.01 & 0.03 & 722 &  52 & 0.14 & 2 \\
pff\_gc-002 & 13:23:59.51 & $-$43:17:29.1 &  22.96 & $ -22.660$ & $   3.728$ & 19.55 &  0.27 &  0.88 &  0.53 &  1.11 & 0.07 & 0.02 & 0.01 & 0.01 & 0.03 & 623 &  52 & 0.14 & 2 \\
pff\_gc-003 & 13:24:03.23 & $-$43:28:13.9 &  31.19 & $ -31.047$ & $  -3.025$ & 19.31 &  0.21 &  0.87 &  0.51 &  1.09 & 0.15 & 0.02 & 0.01 & 0.01 & 0.03 & 697 &  23 & 0.13 & 3 \\
pff\_gc-004 & 13:24:03.74 & $-$43:35:53.4 &  38.00 & $ -37.248$ & $  -7.519$ & 20.01 &  0.28 &  0.83 &  0.52 &  1.09 & 0.28 & 0.03 & 0.01 & 0.01 & 0.03 & 546 &  45 & 0.13 & 3 \\
pff\_gc-005 & 13:24:18.92 & $-$43:14:30.1 &  18.36 & $ -18.185$ & $   2.563$ & 20.29 &  0.58 &  1.02 &  0.58 &  1.22 & 0.18 & 0.04 & 0.01 & 0.01 & 0.03 & 756 &  34 & 0.14 & 3 \\
pff\_gc-006 & 13:24:23.72 & $-$43:07:52.1 &  13.51 & $ -12.260$ & $   5.671$ & 19.22 &  0.21 &  0.78 &  0.52 &  1.02 & 0.05 & 0.02 & 0.01 & 0.01 & 0.03 & 641 &  41 & 0.13 & 2 \\
pff\_gc-007 & 13:24:24.15 & $-$42:54:20.6 &  13.45 & $  -1.160$ & $  13.399$ & 20.12 &  0.69 &  1.04 &  0.59 &  1.30 & 0.17 & 0.03 & 0.01 & 0.01 & 0.03 & 612 &  28 & 0.13 & 3 \\
pff\_gc-008 & 13:24:29.20 & $-$43:21:56.5 &  23.42 & $ -23.188$ & $  -3.252$ & 19.94 &  0.47 &  0.99 &  0.54 &  1.19 & 0.35 & 0.03 & 0.01 & 0.01 & 0.03 & 451 &  43 & 0.14 & 3 \\
pff\_gc-009 & 13:24:31.35 & $-$43:11:26.7 &  14.59 & $ -14.382$ & $   2.473$ & 19.77 &  0.23 &  0.81 &  0.52 &  1.07 & 0.07 & 0.02 & 0.01 & 0.01 & 0.03 & 657 &  46 & 0.13 & 2 \\
pff\_gc-010 & 13:24:33.09 & $-$43:18:44.8 &  20.27 & $ -20.169$ & $  -1.991$ & 19.68 &  0.09 &  0.68 &  0.42 &  0.85 & 0.06 & 0.02 & 0.01 & 0.01 & 0.03 & 344 &  58 & 0.13 & 2 \\
\enddata
\label{table:gcv}
\tablecomments{There are 215 confirmed GCs, of which the first ten are
listed here.  The complete version of this table is in the electronic
edition of the Journal.  The printed edition contains only a sample.}
\end{deluxetable}
}
\end{turnpage}

\clearpage

\begin{turnpage}
{\centering
\begin{deluxetable}{lcccrrrrcccccrrc}
\tabletypesize{\scriptsize}
\tablecaption{Catalog of Confirmed Foreground Stars in Vicinity of \cena}
\tablewidth{0pt}
\tablehead{
\colhead{name} & \colhead{RA} & \colhead{Dec} &
\colhead{$V$} & \colhead{$U$-$B$} & \colhead{$B$-$V$} &
\colhead{$V$-$R$} & \colhead{$V$-$I$} &
\colhead{$\sigma_U$} & \colhead{$\sigma_B$} & 
\colhead{$\sigma_V$} &
\colhead{$\sigma_R$} &\colhead{$\sigma_I$} &
\colhead{$V_{h}$} &
\colhead{$V_{h,err}$} &
\colhead{Run} \\
\colhead{} & \colhead{(J2000)} & \colhead{(J2000)} &
\colhead{mag} & \colhead{mag} & \colhead{mag} &
\colhead{mag} & \colhead{mag} &
\colhead{mag} & \colhead{mag} & 
\colhead{mag} &
\colhead{mag} &\colhead{mag} &
\colhead{km/s} &
\colhead{km/s} &
\colhead{}
}
\startdata
pff\_fs-001 & 13:23:50.43 & $-$43:27:28.3 & 18.05 &  0.17 &  0.70 &  0.41 &  0.83 & 0.06 & 0.01 & 0.01 & 0.01 & 0.03 & $-93$ & $ 18$ & 2 \\
pff\_fs-002 & 13:23:50.74 & $-$43:35:02.6 & 19.28 &  0.71 &  1.10 &  0.65 &  1.47 & 0.32 & 0.02 & 0.01 & 0.01 & 0.03 & $-51$ & $ 31$ & 23 \\
pff\_fs-003 & 13:23:51.83 & $-$43:37:00.1 & 19.01 &  0.17 &  0.71 &  0.42 &  0.91 & 0.12 & 0.02 & 0.01 & 0.01 & 0.03 & $-44$ & $ 31$ & 3 \\
pff\_fs-004 & 13:23:52.56 & $-$42:42:33.5 & 19.72 &  0.19 &  0.67 &  0.68 &  1.43 & 0.10 & 0.03 & 0.01 & 0.01 & 0.03 & $ 71$ & $ 28$ & 3 \\
pff\_fs-005 & 13:23:54.36 & $-$42:48:17.7 & 19.42 & $-$0.09 &  0.70 &  0.34 &  0.92 & 0.05 & 0.02 & 0.01 & 0.01 & 0.03 & $200$ & $ 48$ & 3 \\
pff\_fs-006 & 13:23:55.77 & $-$43:13:15.2 & 20.15 &  0.99 &  1.23 &  0.76 &  1.52 & 0.24 & 0.04 & 0.01 & 0.01 & 0.03 & $-59$ & $ 26$ & 3 \\
pff\_fs-007 & 13:24:02.70 & $-$42:49:54.3 & 20.04 &  0.35 &  0.83 &  0.49 &  1.05 & 0.10 & 0.03 & 0.01 & 0.01 & 0.03 & $-84$ & $ 24$ & 3 \\
pff\_fs-008 & 13:24:02.76 & $-$43:30:15.5 & 18.18 &  0.24 &  0.83 &  0.53 &  1.13 & 0.07 & 0.01 & 0.01 & 0.01 & 0.03 & $-66$ & $ 18$ & 2 \\
pff\_fs-009 & 13:24:03.70 & $-$43:12:13.6 & 20.32 &  1.33 &  1.27 &  0.69 &  1.49 & 0.39 & 0.05 & 0.01 & 0.01 & 0.03 & $-45$ & $ 23$ & 3 \\
pff\_fs-010 & 13:24:03.81 & $-$43:28:05.9 & 19.40 &  1.19 &  1.23 &  0.72 &  1.42 & 0.52 & 0.02 & 0.01 & 0.01 & 0.03 & $ -3$ & $ 24$ & 2 \\
\enddata
\label{table:fgs}
\tablecomments{There are 222 confirmed foreground stars, 
of which the first ten are
listed here.  The complete version of this table is in the electronic
edition of the Journal.  The printed edition contains only a sample.}
\end{deluxetable}
}
\end{turnpage}

\clearpage

\begin{turnpage}
{\centering
\begin{deluxetable}{lcccrrrrccccccc}
\tabletypesize{\scriptsize}
\tablecaption{Catalog of Confirmed Background Galaxies in Vicinity of \cena}
\tablewidth{0pt}
\tablehead{
\colhead{name} & \colhead{RA} & \colhead{Dec} &
\colhead{$V$} & \colhead{\ub} & \colhead{\bv} &
\colhead{\vr} & \colhead{\vi} &
\colhead{$\sigma_U$} & \colhead{$\sigma_B$} & 
\colhead{$\sigma_V$} &
\colhead{$\sigma_R$} &\colhead{$\sigma_I$} &
\colhead{$z$} &
\colhead{Run} \\
\colhead{} & \colhead{(J2000)} & \colhead{(J2000)} &
\colhead{mag} & \colhead{mag} & \colhead{mag} &
\colhead{mag} & \colhead{mag} &
\colhead{mag} & \colhead{mag} & 
\colhead{mag} &
\colhead{mag} &\colhead{mag} &
\colhead{} &
\colhead{}
}
\startdata
pff\_galx-001 & 13:24:44.24 & $-$43:14:36.2 &  20.42 &   0.32 &   1.48 &   0.64 &   1.35 & 0.87 & 0.08 & 0.02 & 0.01 & 0.03 & 0.19 & 3 \\
pff\_galx-002 & 13:25:29.61 & $-$42:29:28.0 &  19.82 &   0.54 &   1.52 &   0.69 &   1.47 & 0.52 & 0.04 & 0.01 & 0.01 & 0.03 & 0.15 & 3 \\
pff\_galx-003 & 13:26:15.68 & $-$42:37:36.0 &  19.91 &   0.08 &   0.53 &   0.39 &   0.77 & 0.18 & 0.02 & 0.01 & 0.01 & 0.03 & 0.06 & 2 \\
pff\_galx-004 & 13:26:29.07 & $-$42:31:38.0 &  18.85 &   0.44 &   0.92 &   0.54 &   1.00 & 0.14 & 0.02 & 0.01 & 0.01 & 0.03 & 0.01 & 1 \\
pff\_galx-005 & 13:26:35.24 & $-$42:34:32.8 &  19.47 &   0.61 &   1.60 &   0.73 &   1.44 & 0.46 & 0.04 & 0.01 & 0.01 & 0.03 & 0.21 & 3 \\
pff\_galx-006 & 13:26:46.04 & $-$42:55:07.9 &  19.09 &   0.23 &   1.00 &   0.64 &   1.26 & 0.05 & 0.02 & 0.01 & 0.01 & 0.03 & 0.10 & 2 \\
pff\_galx-007 & 13:26:47.88 & $-$42:35:36.6 &  20.14 &   0.15 &   0.92 &   0.60 &   1.33 & 0.33 & 0.04 & 0.01 & 0.01 & 0.03 & 0.16 & 3 \\
pff\_galx-008 & 13:26:59.27 & $-$42:17:53.9 &  20.26 &  $-$0.02 &   0.57 &   0.33 &   0.70 & 0.21 & 0.03 & 0.01 & 0.01 & 0.03 & 0.04 & 2 \\
pff\_galx-009 & 13:27:08.48 & $-$42:25:40.6 &  19.48 &   0.66 &   1.80 &   0.77 &   1.57 & 0.64 & 0.05 & 0.01 & 0.01 & 0.03 & 0.22 & 3 \\
pff\_galx-010 & 13:27:09.92 & $-$42:57:08.7 &  18.98 &   0.16 &   0.74 &   0.52 &   1.09 & 0.19 & 0.06 & 0.02 & 0.02 & 0.03 & 0.07 & 2 \\
pff\_galx-011 & 13:27:24.01 & $-$42:43:26.9 &  18.68 &   0.55 &   1.35 &   0.69 &   1.40 & 0.18 & 0.02 & 0.01 & 0.01 & 0.03 & 0.09 & 2 \\
pff\_galx-012 & 13:27:55.02 & $-$42:09:21.0 &  19.28 &   0.27 &   0.76 &   0.50 &   0.61 & 0.21 & 0.02 & 0.01 & 0.01 & 0.03 & 0.02 & 1 \\
pff\_galx-013 & 13:27:55.94 & $-$42:12:46.9 &  19.90 &   0.45 &   1.43 &   0.64 &   1.37 & 0.46 & 0.04 & 0.01 & 0.01 & 0.03 & 0.15 & 3 \\
pff\_galx-014 & 13:28:04.56 & $-$42:13:14.9 &  19.69 &   0.11 &   0.82 &   0.56 &   1.14 & 0.18 & 0.02 & 0.01 & 0.01 & 0.03 & 0.13 & 2 \\
\enddata
\label{table:galx}
\end{deluxetable}
}
\end{turnpage}

\clearpage

\LongTables
\begin{deluxetable}{llccrrrrrrrrrr}
\tabletypesize{\scriptsize}
\tablecaption{Optical Matches with X-Ray Point Sources
\label{table:xraymatch}}
\tablewidth{0pt}
\tablehead{
\colhead{KraftID} &
\colhead{Other ID} &
\colhead{RA} &
\colhead{Dec} &
\colhead{$V$} & \colhead{$U$-$B$} & \colhead{$B$-$V$} &
\colhead{$V$-$R$} & \colhead{$V$-$I$} &
\colhead{$\sigma_U$} & \colhead{$\sigma_B$} &
\colhead{$\sigma_V$} &
\colhead{$\sigma_R$} &\colhead{$\sigma_I$} \\
\colhead{} &
\colhead{} &
\colhead{(J2000)} &
\colhead{(J2000)} &
\colhead{mag} &
\colhead{mag} &
\colhead{mag} &
\colhead{mag} &
\colhead{mag} &
\colhead{mag} &
\colhead{mag} &
\colhead{mag} &
\colhead{mag} &
\colhead{mag}
}
\startdata
K-007 & \nodata & 13:24:58.17 & $-$43:09:49.19 &  20.54 &  $-$0.53 &   0.45 &   0.21 &   0.59 &   0.07 &   0.04 &   0.02 &   0.02 &   0.04 \\
K-015 & HGHH-G176    & 13:25:03.12 & $-$42:56:25.07 &  18.93 &   0.54 &   1.01 &   0.61 &   1.26 &   0.09 &   0.02 &   0.01 &   0.01 &   0.03 \\
K-022 & \nodata & 13:25:05.71 & $-$43:10:30.83 &  18.00 &   0.67 &   1.02 &   0.58 &   1.29 &   0.04 &   0.01 &   0.01 &   0.01 &   0.03 \\
K-029 & \nodata & 13:25:09.19 & $-$42:58:59.19 &  17.71 &   0.48 &   0.94 &   0.58 &   1.16 &   0.08 &   0.02 &   0.01 &   0.01 &   0.03 \\
K-030 & \nodata & 13:25:09.17 & $-$42:59:17.84 &  20.97 &  $-$0.32 &   0.25 &   0.11 &   0.32 &   0.37 &   0.17 &   0.08 &   0.10 &   0.15 \\
K-033 & \nodata & 13:25:10.25 & $-$42:55:09.54 &  19.46 &   0.60 &   1.01 &   0.57 &   1.25 &   0.13 &   0.03 &   0.01 &   0.01 &   0.03 \\
K-034 & \nodata & 13:25:10.27 & $-$42:53:33.13 &  17.80 &   0.55 &   0.99 &   0.54 &   1.25 &   0.04 &   0.01 &   0.01 &   0.01 &   0.03 \\
K-037 & \nodata & 13:25:11.03 & $-$42:52:58.23 &  19.93 &   1.07 &   1.48 &   1.33 &   2.93 &   0.40 &   0.05 &   0.01 &   0.01 &   0.03 \\
K-041 & \nodata & 13:25:11.98 & $-$42:57:13.32 &  19.09 &   0.29 &   0.87 &   0.53 &   1.11 &   0.09 &   0.03 &   0.01 &   0.01 &   0.03 \\
K-047 & \nodata & 13:25:13.82 & $-$42:53:31.09 &  20.64 &  $-$0.26 &   1.09 &   0.84 &   1.70 &   0.18 &   0.07 &   0.02 &   0.02 &   0.03 \\
K-051 & \nodata & 13:25:14.25 & $-$43:07:23.60 &  19.55 &   0.64 &   1.07 &   0.65 &   1.34 &   0.18 &   0.04 &   0.01 &   0.01 &   0.03 \\
K-060 & \nodata & 13:25:19.62 & $-$42:49:23.89 &  20.50 &  $-$0.30 &   0.48 &   0.26 &   0.75 &   0.07 &   0.03 &   0.02 &   0.02 &   0.03 \\
K-073 & \nodata & 13:25:22.67 & $-$42:55:01.63 &  21.11 &  $-$0.51 &   0.59 &   0.32 &   0.72 &   0.15 &   0.08 &   0.03 &   0.03 &   0.05 \\
K-102 & \nodata & 13:25:27.98 & $-$43:04:02.22 &  19.18 &   0.66 &   1.06 &   0.61 &   1.28 &   0.56 &   0.11 &   0.03 &   0.02 &   0.04 \\
K-110 & \nodata & 13:25:29.10 & $-$43:07:46.22 &  21.23 &   0.83 &   1.32 &   0.72 &   1.50 &   1.12 &   0.16 &   0.04 &   0.03 &   0.04 \\
K-116 & \nodata & 13:25:31.05 & $-$43:11:07.09 &  19.86 &  $-$0.60 &   0.16 &   0.39 &   0.67 &   0.03 &   0.02 &   0.01 &   0.01 &   0.03 \\
K-122 & \nodata & 13:25:32.16 & $-$43:10:40.95 &  19.21 &   0.79 &   0.97 &   0.58 &   1.06 &   0.11 &   0.02 &   0.01 &   0.01 &   0.03 \\
K-127 & HGHH-G359    & 13:25:32.42 & $-$42:58:50.17 &  18.86 &   0.58 &   1.00 &   0.62 &   1.22 &   0.45 &   0.10 &   0.03 &   0.02 &   0.04 \\
K-130 & pff\_gc-056 & 13:25:32.80 & $-$42:56:24.43 &  18.64 &   0.19 &   0.79 &   0.48 &   1.00 &   0.06 &   0.02 &   0.01 &   0.01 &   0.03 \\
K-131 & \nodata & 13:25:32.88 & $-$43:04:29.19 &  19.37 &   0.79 &   1.08 &   0.59 &   1.24 &   0.38 &   0.07 &   0.02 &   0.02 &   0.03 \\
K-137 & \nodata & 13:25:33.58 & $-$43:12:40.42 &  19.75 &  $-$0.17 &   0.51 &   0.49 &   1.13 &   0.05 &   0.02 &   0.01 &   0.01 &   0.03 \\
K-140 & HGHH-G206    & 13:25:34.10 & $-$42:59:00.68 &  19.10 &   0.58 &   0.99 &   0.52 &   1.19 &   0.56 &   0.12 &   0.03 &   0.03 &   0.04 \\
K-141 & \nodata & 13:25:34.05 & $-$43:10:31.13 &  20.10 &  $-$0.13 &   1.19 &   0.58 &   1.33 &   0.12 &   0.05 &   0.01 &   0.01 &   0.03 \\
K-144 & \nodata & 13:25:35.16 & $-$42:53:00.96 &  20.25 &   0.79 &   1.06 &   0.62 &   1.31 &   0.33 &   0.05 &   0.02 &   0.01 &   0.03 \\
K-147 & \nodata & 13:25:35.50 & $-$42:59:35.20 &  19.41 &   0.66 &   1.05 &   0.62 &   1.32 &   1.08 &   0.20 &   0.05 &   0.04 &   0.05 \\
K-156 & HGHH-G268    & 13:25:38.61 & $-$42:59:19.52 &  18.93 &   0.40 &   0.91 &   0.56 &   1.13 &   0.34 &   0.08 &   0.03 &   0.02 &   0.04 \\
K-159 & \nodata & 13:25:39.08 & $-$42:56:53.64 &  19.90 &  $-$0.24 &   0.45 &   0.26 &   0.64 &   0.09 &   0.04 &   0.02 &   0.02 &   0.04 \\
K-163 & HHH86-18     & 13:25:39.88 & $-$43:05:01.91 &  17.53 &   0.38 &   0.89 &   0.56 &   1.10 &   0.04 &   0.01 &   0.01 &   0.01 &   0.03 \\
K-170 & R202         & 13:25:42.00 & $-$43:10:42.21 &  19.26 &   0.27 &   0.81 &   0.54 &   1.05 &   0.06 &   0.02 &   0.01 &   0.01 &   0.03 \\
K-172 & pff\_gc-062 & 13:25:43.23 & $-$42:58:37.39 &  19.42 &   0.67 &   1.04 &   0.59 &   1.24 &   0.34 &   0.06 &   0.02 &   0.02 &   0.03 \\
K-182 & \nodata & 13:25:46.34 & $-$43:03:10.22 &  21.34 &   0.04 &   1.14 &   0.75 &   1.69 &   0.82 &   0.26 &   0.06 &   0.04 &   0.04 \\
K-184 & HGHH-G284    & 13:25:46.59 & $-$42:57:02.97 &  19.87 &   0.57 &   1.03 &   0.58 &   1.24 &   0.26 &   0.05 &   0.02 &   0.01 &   0.03 \\
K-192 & \nodata & 13:25:48.71 & $-$43:03:23.41 &  18.72 &   1.27 &   1.24 &   0.69 &   1.31 &   0.25 &   0.03 &   0.01 &   0.01 &   0.03 \\
K-197 & \nodata & 13:25:49.86 & $-$42:51:18.27 &  20.95 &   0.79 &   1.73 &   0.85 &   1.70 &   0.90 &   0.13 &   0.02 &   0.02 &   0.03 \\
K-199 & HGHH-21      & 13:25:52.74 & $-$43:05:46.54 &  17.87 &   0.40 &   0.89 &   0.55 &   1.11 &   0.04 &   0.01 &   0.01 &   0.01 &   0.03 \\
K-200 & \nodata & 13:25:53.63 & $-$43:01:32.98 &  19.01 &   1.00 &   1.43 &   0.86 &   1.68 &   0.39 &   0.05 &   0.01 &   0.01 &   0.03 \\
K-201 & \nodata & 13:25:53.75 & $-$43:11:55.60 &  20.62 &  $-$0.05 &   1.23 &   0.81 &   1.58 &   0.18 &   0.06 &   0.02 &   0.01 &   0.03 \\
K-202 & HGHH-23      & 13:25:54.59 & $-$42:59:25.37 &  17.22 &   0.63 &   1.07 &   0.60 &   1.28 &   0.04 &   0.01 &   0.01 &   0.01 &   0.03 \\
K-204 & \nodata & 13:25:55.13 & $-$43:01:18.29 &  21.37 &   0.33 &   0.49 &   0.34 &   0.64 &   0.64 &   0.16 &   0.06 &   0.07 &   0.10 \\
K-207 & \nodata & 13:25:56.87 & $-$43:00:44.40 &  20.58 &  $-$0.60 &   0.46 &   0.37 &   0.95 &   0.10 &   0.06 &   0.03 &   0.03 &   0.04 \\
K-209 & \nodata & 13:25:57.42 & $-$42:53:41.60 &  19.14 &  $-$0.46 &   0.31 &   0.44 &   0.80 &   0.03 &   0.02 &   0.01 &   0.01 &   0.03 \\
K-213 & \nodata & 13:25:57.42 & $-$42:53:41.60 &  19.14 &  $-$0.46 &   0.31 &   0.44 &   0.80 &   0.03 &   0.02 &   0.01 &   0.01 &   0.03 \\
K-216 & \nodata & 13:25:58.71 & $-$43:04:30.71 &  19.02 &  $-$0.25 &   0.24 &   0.20 &   0.53 &   0.03 &   0.02 &   0.01 &   0.01 &   0.03 \\
K-217 & \nodata & 13:26:00.81 & $-$43:09:40.07 &  20.09 &   0.54 &   0.99 &   0.56 &   1.19 &   0.16 &   0.03 &   0.01 &   0.01 &   0.03 \\
K-218 & \nodata & 13:26:01.12 & $-$43:05:29.24 &  20.89 &   1.00 &   1.62 &   1.21 &   2.67 &   1.07 &   0.13 &   0.02 &   0.01 &   0.03 \\
K-220 & HGHH-07      & 13:26:05.41 & $-$42:56:32.38 &  17.17 &   0.33 &   0.87 &   0.54 &   1.08 &   0.03 &   0.01 &   0.01 &   0.01 &   0.03 \\
K-223 & HGHH-37      & 13:26:10.58 & $-$42:53:42.68 &  18.43 &   0.48 &   0.95 &   0.56 &   1.17 &   0.04 &   0.01 &   0.01 &   0.01 &   0.03 \\
K-224 & \nodata & 13:26:11.86 & $-$43:02:43.24 &  21.18 &  $-$0.30 &   0.38 &   0.52 &   1.16 &   0.13 &   0.06 &   0.03 &   0.02 &   0.04 \\
K-230 & \nodata & 13:26:16.09 & $-$42:58:45.57 &  20.90 &   1.29 &   1.54 &   0.90 &   1.75 &   1.30 &   0.12 &   0.02 &   0.02 &   0.03 \\
K-232 & \nodata & 13:26:16.09 & $-$42:58:45.57 &  20.90 &   1.29 &   1.54 &   0.90 &   1.75 &   1.30 &   0.12 &   0.02 &   0.02 &   0.03 \\
K-233 & \nodata & 13:26:19.66 & $-$43:03:18.64 &  18.74 &   0.41 &   0.93 &   0.58 &   1.17 &   0.05 &   0.02 &   0.01 &   0.01 &   0.03 \\
K-235 & \nodata & 13:26:20.42 & $-$42:59:46.35 &  21.20 &  $-$0.50 &   0.40 &   0.23 &   0.75 &   0.10 &   0.05 &   0.02 &   0.03 &   0.04 \\
\enddata
\end{deluxetable}

\clearpage

\begin{deluxetable}{lcccrrrrc}
\tabletypesize{\scriptsize}
\tablecaption{White Dwarfs and QSOs in the Field of \cena
\label{table:blueobj}}
\tablewidth{0pt}
\tablehead{
\colhead{ID} & \colhead{RA(J2000)} & \colhead{DEC(J2000)} &
\colhead{V} & \colhead{\ub} & \colhead{\bv} & \colhead{\vr} &
\colhead{\vi} & \colhead{Redshift}
}
\startdata
pff\_wd-1 & 13:26:06.70 & $-$42:39:52.5 & 19.13 & $-$0.26 & 0.15 & $-$0.02 & 
$-$0.07 & \nodata \\
pff\_wd-2 & 13:23:59.46 & $-$43:20:43.6 & 19.75 & $-$0.34 & 0.26 & 0.06 & 
0.23 & \nodata \\
pff\_wd-3 & 13:26:47.83 & $-$43:33:42.9 & 19.11 & $-$0.35 & 0.02 & $-$0.06 & 
$-$0.18 & \nodata \\
pff\_qso-1 & 13:25:06.46 & $-$42:29:02.7 & 19.64 & $-$0.18 & 0.29 & 0.31 & 
0.75 & 0.7 \\
pff\_qso-2 & 13:27:55.97 & $-$42:42:32.9 & 19.83 & $-$0.33 & 0.29 & 0.22 & 
0.58 & 0.7 \\
pff\_qso-3 & 13:24:43.45 & $-$43:27:12.0 & 18.59 & $-$0.71 & 0.20 & 0.29 & 
0.68 & 0.7 \\
pff\_qso-4 & 13:25:38.64 & $-$43:25:32.2 & 19.68 & $-$0.25 & 0.37 & 0.28 & 
0.80 & 0.8 \\
pff\_qso-5 & 13:27:36.08 & $-$42:14:32.3 & 19.99 & $-$0.15 & 0.55 & 0.38 & 
0.98 & 0.8 \\
pff\_qso-6 & 13:25:31.05 & $-$43:11:07.0 & 19.86 & $-$0.60 & 0.16 & 0.39 & 
0.67 & 0.6 \\
pff\_qso-7 & 13:25:55.83 & $-$43:14:42.1 & 19.58 & $-$0.52 & 0.30 & 0.27 & 
0.68 & 0.7 \\
\enddata
\end{deluxetable}



\begin{thebibliography}{}
\bibitem[Ashman \& Zepf(1992)]{1992ApJ...384...50A} Ashman, K.~M.~\& Zepf, S.~E.\ 1992, \apj, 384, 50 
\bibitem[Beasley et al.(2002)]{2002MNRAS.333..383B} Beasley, M.~A., Baugh, C.~M., Forbes, D.~A., Sharples, R.~M., \& Frenk, C.~S.\ 2002, \mnras, 333, 383 
\bibitem[Bertin \& Arnouts(1996)]{1996A&AS..117..393B} Bertin, E.~\& 
Arnouts, S.\ 1996, \aaps, 117, 393 
\bibitem[Canterna(1976)]{1976AJ.....81..228C} Canterna, R.\ 1976, \aj, 81, 
228 
\bibitem[C{\^o}t{\'e}, Marzke, \& West(1998)]{1998ApJ...501..554C} 
C{\^o}t{\'e}, P., Marzke, R.~O., \& West, M.~J.\ 1998, \apj, 501, 554 
\bibitem[Dekel \& Silk(1986)]{1986ApJ...303...39D} Dekel, A.~\& Silk, J.\ 
1986, \apj, 303, 39 
\bibitem[Ebneter \& Balick(1983)]{1983PASP...95..675E} Ebneter, K.~\& 
Balick, B.\ 1983, \pasp, 95, 675 
\bibitem[Forbes, Brodie, \& Grillmair(1997)]{1997AJ....113.1652F} Forbes, 
D.~A., Brodie, J.~P., \& Grillmair, C.~J.\ 1997, \aj, 113, 1652 
\bibitem[Graham(1998)]{1998ApJ...502..245G} Graham, J.~A.\ 1998, \apj, 502, 
245 
\bibitem[Graham \& Phillips(1980)]{1980ApJ...239L..97G} Graham, J.~A.~\& 
Phillips, M.~M.\ 1980, \apjl, 239, L97 [GP80]
\bibitem[Harris, Hesser, Harris, \& Curry(1984)]{1984ApJ...287..175H} 
Harris, G.~L.~H., Hesser, J.~E., Harris, H.~C., \& Curry, P.~J.\ 1984, 
\apj, 287, 175 
\bibitem[Harris, Geisler, Harris, \& Hesser(1992)]{1992AJ....104..613H} 
Harris, G.~L.~H., Geisler, D., Harris, H.~C., \& Hesser, J.~E.\ 1992, \aj, 
104, 613 [HGHH92]
\bibitem[Harris \& Harris(2000)]{2000AJ....120.2423H} Harris, G.~L.~H.~\& 
Harris, W.~E.\ 2000, \aj, 120, 2423 [HH00]
\bibitem[Harris \& Harris(2000)]{2003...submitted} Harris, G.~L.~H.~\& 
Harris, W.~E.\ 2003, submitted
\bibitem[Harris, Harris, Holland, \& McLaughlin(2002)]{2002AJ....124.1435H} 
Harris, W.~E., Harris, G.~L.~H., Holland, S.~T., \& McLaughlin, D.~E.\ 
2002, \aj, 124, 1435
\bibitem[Held, Federici, Testa, \& Cacciari(1997)]{1997neg..conf..500H} 
Held, E.~V., Federici, L., Testa, V., \& Cacciari, C.\ 1997, ASP 
Conf.~Ser.~116: The Nature of Elliptical Galaxies; 2nd Stromlo Symposium, 
500 
\bibitem[Hesser, Harris, \& Harris(1986)]{1986ApJ...303L..51H} Hesser, 
J.~E., Harris, H.~C., \& Harris, G.~L.~H.\ 1986, \apjl, 303, L51 [HHH86]
\bibitem[Hui, Ford, Ciardullo, \& Jacoby(1993)]{hui93} Hui, X., Ford, H.~C., Ciardullo, R., \& Jacoby, G.~H.\ 1993, \apj, 414, 463
\bibitem[Holland, C{\^ o}t{\' e}, \& Hesser(1999)]{1999A&A...348..418H} 
Holland, S., C{\^ o}t{\' e}, P., \& Hesser, J.~E.\ 1999, \aap, 348, 418
[HCH99] 
\bibitem[Israel(1998)]{1998A&ARv...8..237I} Israel, F.~P.\ 1998, \aapr, 8, 
237 
\bibitem[Kundu \& Whitmore(2001)]{2001AJ....121.2950K} Kundu, A.~\& 
Whitmore, B.~C.\ 2001, \aj, 121, 2950 
\bibitem[Kundu, Maccarone, \& Zepf(2002)]{2002ApJ...574L...5K} Kundu, A., 
Maccarone, T.~J., \& Zepf, S.~E.\ 2002, \apjl, 574, L5 
\bibitem[Landolt(1992)]{1992AJ....104..340L} Landolt, A.~U.\ 1992, \aj, 
104, 340 
\bibitem[Larsen et al.(2001)]{2001AJ....121.2974L} Larsen, S.~S., Brodie, 
J.~P., Huchra, J.~P., Forbes, D.~A., \& Grillmair, C.~J.\ 2001, \aj, 121, 
2974 
\bibitem[Malin(1978)]{1978Natur.276..591M} Malin, D.~F.\ 1978, \nat, 276, 
591 
\bibitem[Malin \& Carter(1983)]{1983ApJ...274..534M} Malin, D.~F.~\& 
Carter, D.\ 1983, \apj, 274, 534 
\bibitem[Malin \& Hadley(1997)]{mh97} Malin, D. \& Hadley, B. 1997, in ``The Nature of Elliptical Galaxies; 2nd Stromlo Symposium. ASP Conference Series; Vol. 116; 1997; ed. M. Arnaboldi; G. S. Da Costa; and P. Saha (1997), p.460
\bibitem[Minniti et al.(1996)]{1996ApJ...467..221M} Minniti, D., Alonso, 
M.~V., Goudfrooij, P., Jablonka, P., \& Meylan, G.\ 1996, \apj, 467, 221 
[MAGJM96]
\bibitem[Minniti \& Rejkuba(2002)]{2002ApJ...575L..59M} Minniti, D.~\& 
Rejkuba, M.\ 2002, \apjl, 575, L59 [MR02]
\bibitem[Minniti, Rejkuba, Funes, \& Akiyama(2003)]{2003ApJ...Minniti}
Minniti, D., Rejkuba, M., Funes, J.~G., \& Akiyama, S.\ 2003, \apj, submitted
\bibitem[Peng, Ford, Freeman, \& White(2002)]{2002AJ....124.3144P} Peng, 
E.~W., Ford, H.~C., Freeman, K.~C., \& White, R.~L.\ 2002, \aj, 124, 3144 
\bibitem[Peng, Ford, \& Freeman(2004)]{2004...ApJinpress} Peng, 
E.~W., Ford, H.~C., \& Freeman, K.~C.\ 2004, \apj, 602, in press,
astro-ph/0311236
\bibitem[Rejkuba(2001)]{2001A&A...369..812R} Rejkuba, M.\ 2001, \aap, 369, 
812 
\bibitem[Schlegel, Finkbeiner, \& Davis(1998)]{1998ApJ...500..525S} 
Schlegel, D.~J., Finkbeiner, D.~P., \& Davis, M.\ 1998, \apj, 500, 525
\bibitem[Sharples(1988)]{1988IAUS..126..545S} Sharples, R.\ 1988, IAU 
Symposium, 126, 545 
\bibitem[Skuljan, Hearnshaw, \& Cottrell(2000)]{2000PASP..112..966S} 
Skuljan, J., Hearnshaw, J.~B., \& Cottrell, P.~L.\ 2000, \pasp, 112, 966 
\bibitem[van den Bergh, Hesser, \& Harris(1981)]{1981AJ.....86...24V} van 
den Bergh, S., Hesser, J.~E., \& Harris, G.~L.~H.\ 1981, \aj, 86, 24 [VHH81]
\bibitem[Zepf \& Ashman(1993)]{1993MNRAS.264..611Z} Zepf, S.~E.~\& Ashman, 
K.~M.\ 1993, \mnras, 264, 611 
\bibitem[Zickgraf, Humphreys, Graham, \& 
Phillips(1990)]{1990PASP..102..920Z} Zickgraf, F., Humphreys, R.~M., 
Graham, J.~A., \& Phillips, A.\ 1990, \pasp, 102, 920 
\bibitem[Zirm, Dickinson, \& Dey(2003)]{2003ApJ...585...90Z} Zirm, A.~W., 
Dickinson, M., \& Dey, A.\ 2003, \apj, 585, 90 
\end{thebibliography}
\end{document}